\documentclass[a4paper,fleqn]{cas-sc}

\usepackage{natbib}
\usepackage{amssymb}
\usepackage{amsmath}
\usepackage{tabularx}
\usepackage{cancel}
\usepackage{multirow}
\usepackage{array} 
\usepackage{booktabs}
\usepackage{supertabular}
\usepackage{algorithm}
\usepackage{algorithmic}
\usepackage[shortlabels]{enumitem}
\usepackage{amsthm}
\usepackage{float}
\usepackage{subfig}
\usepackage{graphicx}
\usepackage{rotating}
\usepackage{setspace}
\usepackage{subcaption}

\usepackage{xcolor}
\numberwithin{equation}{section}

\usepackage{ulem}
\usepackage{multirow}

\def\tsc#1{\csdef{#1}{\textsc{\lowercase{#1}}\xspace}}
\tsc{WGM}
\tsc{QE}

\newcommand{\unit}[1]{\mathrm{#1}}

\newcommand{\R}{\mathbb{R}}  

\newcommand{\Esp}{\mathbb{E}}  

\newcommand{\Var}{\mathbb{V}\mathrm{ar}}  
\newcommand{\indexSet}{I}   
\newcommand{\exceptIndexSet}{- \indexSet}  

\newcommand{\nvar}{d}   
\newcommand{\nOutputDimensions}{L}  
\newcommand{\indPoints}{i}

\newcommand{\nDoE}{n}  

\newcommand{\SobolNormalized}[1]{S_{#1}}  
\newcommand{\SobolUnnormalized}[1]{D_{#1}}  
\newcommand{\sobolTot}[1]{\SobolNormalized{#1}^{\text{T}}}
\newcommand{\sobolTotD}[1]{\SobolUnnormalized{#1}^{\text{T}}}
\newcommand{\sobolClosed}[1]{\SobolNormalized{#1}^{\text{c}}}  
\newcommand{\sobolClosedD}[1]{\SobolUnnormalized{#1}^{\text{c}}}  
\newcommand{\totalVariance}{D}

\newcommand{\indOutputs}{\ell}
\newcommand{\inputVectorRandom}{X}  
\newcommand{\inputPointRandom}[1]{\inputVectorRandom_{#1}}
\newcommand{\deterministicModel}{f}  
\newcommand{\inputPoint}{x}  
\newcommand{\DoE}{\mathcal{X}}   

\newcommand{\pfEstim}[1]{{\widehat{#1}}^{\mathrm{pf}}}  

\newcommand{\indGPR}{j}
\newcommand{\GaussianProcess}{Z} 
\newcommand{\MeanFunction}{m}
\newcommand{\KernelFunction}{K}
\newcommand{\condGP}{\GaussianProcess_c}
\newcommand{\ConditionalMean}{\MeanFunction_c}
\newcommand{\ConditionalKernel}{\KernelFunction_c}
\newcommand{\condGPVector}{\GaussianProcess_n}
\newcommand{\ObservationVector}{\outputVectorDeterministic}
\newcommand{\KernelMatrix}{\KernelFunction(\DoE, \DoE)}
\newcommand{\KernelVector}[1]{\KernelFunction(\DoE, #1)}
\newcommand{\KernelVectorRow}[1]{\KernelFunction(#1, \DoE)}
\newcommand{\NumberGPite}{N_{Z}}
\newcommand{\nX}{N_X}

\newcommand{\indBasis}{q}
\newcommand{\coef}{c}   
\newcommand{\coefPred}{\coef^{*}}
\newcommand{\basis}{v}  
\newcommand{\basisVec}{\basis_{., \indOutputs}}   

\newcommand{\nbasis}{n_{b}}  
\newcommand{\GramMatrix}{\mathbf{G}}

\newcommand{\depth}{h}
\newcommand{\depthThresh}{H_f}
\newcommand{\timeSteps}{t_n}
\newcommand{\numberTimeSteps}{N_t}
\newcommand{\velocityVec}{\vec{V}}
\newcommand{\velocityX}{u}
\newcommand{\velocityY}{v}
\newcommand{\gradientVec}{\vec{\nabla}}
\newcommand{\gravity}{g}
\newcommand{\gravityVec}{\vec{\gravity}}
\newcommand{\shearStress}{\tau_b}
\newcommand{\densityMixture}{\rho_m}
\newcommand{\hydraulicRad}{R}
\newcommand{\manning}{\text{n}}
\newcommand{\yieldStress}{\tau_c}
\newcommand{\plasticViscosity}{\mu_B}
\newcommand{\shearRate}{\dot{\gamma}}
\newcommand{\relativeTimePeak}{r_p}
\newcommand{\totalTimeHyd}{t_T}
\newcommand{\densityWater}{\rho_w}
\newcommand{\densitySolid}{\rho_s}
\newcommand{\volumetricConc}{C_v}
\newcommand{\releasedVol}{Vol}
\newcommand{\arrivalTime}{T}

\newcommand{\maxFlowDepth}{H}

\newcommand{\maxVel}{V}

\begin{document}
\let\WriteBookmarks\relax
\def\floatpagepagefraction{1}
\def\textpagefraction{.001}

\shorttitle{Probabilistic analysis of the Brumadinho tailings dam-breach}    

\shortauthors{Y.T. Sáo}  

\title [mode = title]{A probabilistic analysis of the Brumadinho tailings dam-breach using uncertainty quantification and global sensitivity analysis}

%

\author[1,2]{Yuri Taglieri Sáo}[orcid=0000-0003-2037-5451]
\cormark[1]

\ead{yuri.sao@unesp.br}

\credit{Conceptualization, methodology, investigation, software, validation, formal analysis, visualization, writing (original draft), writing (review and editing)}

\affiliation[1]{organization={Department of Mechanical Engineering, São Paulo State University ``Júlio de Mesquita Filho'' (UNESP)},
            addressline={Av. Brasil, 56}, 
            city={Ilha Solteira},
            postcode={15385-000}, 
            state={São Paulo},
            country={Brazil}}

\affiliation[2]{organization={Department of Mathematics and Applications, Institut National des Sciences Appliquées (INSA)},
            addressline={135, Avenue de Rangueil}, 
            city={Toulouse},
            postcode={31077}, 
            state={Occitanie},
            country={France}}

\author[1]{Geraldo de Freitas Maciel}

\credit{Conceptualization, writing (review and editing), supervision, funding acquisition}

\author[3]{Julian Cardoso Eleutério}

\credit{Data curation, investigation, writing (review and editing)}

\affiliation[3]{organization={Department of Hydraulic and Water Resources Engineering, Federal University of Minas Gerais (UFMG)},
            addressline={Escola de Engenharia}, 
            city={Belo Horizonte},
            postcode={31270-901}, 
            state={Minas Gerais},
            country={Brazil}}

\cortext[1]{Yuri Taglieri Sáo}


\begin{abstract}
On 25th January 2019, Dam B1 in Brumadinho, Brazil, suddenly collapsed and produced a catastrophic tailings flow with severe human, environmental and infrastructural consequences. Previous numerical back-analyses of the tailings flow have relied on deterministic approaches, which provide useful reference results but do not explicitly incorporate uncertainty associated to rupture hydrograph and tailings' physical properties. This work presents a probabilistic approach to study the tailings dam-breach flow caused by the rupture of Dam B1, focusing on the quantification and interpretation of the effects of input uncertainty on hazard-related hydrodynamical outputs. Deterministic data was obtained through HEC-RAS v6.6 by considering tailings as an one-phase Bingham fluid. The probabilistic framework combines dimensionality reduction, metamodeling and flooding classification to perform fast prediction of unexplored input scenarios. It allows to obtain data distributions that reflect output uncertainty and to estimate variance-based sensitivity indices to quantify the influence of input uncertainty over output variability. Results show that uncertainty is strongly dependent on time and spatial location. Hydrograph-related parameters mainly control the timing and magnitude of the early flow, whereas yield stress becomes more influential in regions associated with flow attenuation, spreading and possible arrest. Overall, the framework provides probabilistic hazard estimates and insights into the physical mechanisms governing the dynamics of tailings flow. 
\end{abstract}




\maketitle

\section{Introduction}

On 25th January 2019 at 12:28 local time, the tailings dam ``Dam B1'' of the iron-ore mine Córrego do Feijão collapsed near the municipality of Brumadinho, in Brazil. The failure released a large volume of tailings that produced a rapid and deep mudflow, devastating the downstream valley, including the mine's administrative buildings and the Paraopeba river \citep{Rotta2020, Lumbroso2021}. The mudflow killed 270 people, most of them mine workers, and the event is known as the ``Brumadinho dam disaster'', regarded as the worst work-related disaster in Brazil history and one of the worst mining disasters ever reported \citep{Rotta2020}. Beyond the human loss of life, downstream communities had their water supply interrupted and several environmental impacts were observed, such as contamination of the Paraopeba river and damage to the ecosystem \citep{Vergilio2020, Thompson2020, Rotta2020}. In a technical sense, this unprecedented disaster exposes the importance of understanding hydrodynamical/geotechnical properties, or \textit{hazard-related properties}, of the mass movement produced by a tailings-dam failure, since they are directly associated with downstream risks and consequences. Some examples of hazard-related properties are flow depth, velocity, arrival times and inundation extents.

Hazard-related properties are commonly estimated through physical modelling of the tailings dam-breach flow, usually through numerical solvers. In the case of already breached dams, this type of investigation is referred to as \textit{back-analysis} and its general procedure is similar to that adopted in tailings-dam breach analyses (TDBAs) of unbreached dams \citep{Martin2019, Liu2020_dam, Gildeh2021}. Briefly, prior information about the dam and the event, such as pre-rupture conditions, mobilized (or released) volume, possible breach mechanisms, terrain characteristics and physical properties of tailings, is gathered and incorporated into a deterministic model. The model receives inputs correspondent to the prior information and propagates the tailings flow over the complex downstream terrain, providing estimates of the hazard-related properties. For the Brumadinho case, back-analysis is useful to improve the understanding of the tailings flow behaviour throughout its course, providing details and estimates that could not be obtained from observation alone. 

However, tailings dam-breach flows are strongly affected by uncertainties in the aforementioned inputs \citep{Rico2008a, Ghahramani2020, Tian2021, Walsh2021, Moon2019, Piette2022, Ghahramani2024}. These uncertainties are particularly relevant because their effects on hazard-related properties are not known a priori and may be propagated through the model in a non-linear manner, affecting different quantities of interest in different ways over space and time. Although some studies have performed back-analyses of the Brumadinho disaster by exploring one or a few representative scenarios \citep{Yu2020, Ligier2020, Lumbroso2021, Gibson2022, Melo2022}, purely deterministic approaches do not allow a systematic characterization of the range of plausible consequences implied by uncertain inputs. In this sense, probabilistic techniques such as uncertainty quantification (UQ) and global sensitivity analysis (GSA) are particularly useful and complementary. While UQ aims to propagate and characterize uncertainty in the outputs, GSA allows quantifying the influence of input uncertainty on output variability \citep{Soize2017, Saltelli2008}. Therefore, the combination of UQ and GSA is particularly useful to support the physical interpretation of the Brumadinho tailings flow by identifying which uncertain inputs are most influential on hazard-related properties, both at selected control points and over the spatial domain. Similar hydraulics applications were done by \cite{Li2020, Zegers2020, Elgarroussi2022, Rohmer2022}.

In tailings dam-breach literature, probabilistic approaches have received limited attention \citep{Melo2023, Ghahramani2024}. In this context, the present work proposes a systematic probabilistic back-analysis of the Brumadinho dam-breach disaster, shifting the standard deterministic reconstruction to a probabilistic one. The main contribution of the study is the combination of UQ and GSA to investigate high-dimensional spatiotemporal hazard-related properties of the Brumadinho dam-breach flow, while preserving physical interpretability of the results. More specifically, the employed framework makes the probabilistic analysis of the Brumadinho case computationally feasible by dealing with the high dimensionality of the outputs using a dimensionality-reduction technique and by enabling the efficient exploration of unexplored scenarios under uncertain conditions through metamodeling. This allows the characterization of plausible consequences in terms of hazard-related properties (flow depth over time, maximum flow depth, maximum velocity, arrival time and flooding probability) and the identification of the dominant uncertain inputs governing these properties. Therefore, the contributions of the present work are twofold: first, to provide a systematic probabilistic characterization of the Brumadinho tailings flow, both at selected control points and over the spatial domain; and second, to support the physical interpretation of the predicted consequences through aggregated sensitivity measures and spatial Sobol' maps, indicating which inputs are influential, over which outputs, and where in the downstream domain their effects are most relevant.

The remainder of this manuscript is organized as follows. Section 2 presents the case study and the hydraulic modeling strategy. Section 3 describes the uncertain inputs and the probabilistic framework adopted for UQ and GSA. Section 4 presents the main results in terms of uncertainty propagation and sensitivity analysis. Finally, Section 5 summarizes the main conclusions and perspectives.

\section{Case study and deterministic model}
\label{sec:deterministic_model}

\subsection{The Brumadinho dam-breach case}
\label{subsec:brumadinho_description}

Dam B1 of the Córrego do Feijão mine was an inactive tailings dam used for the storage of iron-ore tailings. It had a height of $86 \ \unit{m}$, a crest length of $700 \ \unit{m}$ and stored approximately $11.7 \times 10^6 \ \unit{m}^3$ of tailings. Over a period of 37 years, the dam was raised using the upstream method, in which successive embankment raises are constructed over previously deposited tailings, progressing upstream from an initial starter dam \citep{Blight2009, Robertson2019}. On 25 January 2019, at 12:28 local time, Dam B1 suddenly collapsed.

\begin{figure}[h]
\centering

\subfloat[Downstream area immediately affected by the dam breach. From Felipe Werneck, Instituto Brasileiro do Meio Ambiente e dos Recursos Naturais Renováveis (IBAMA) – 01/02/2019.]{
    \includegraphics[width=0.48\textwidth]{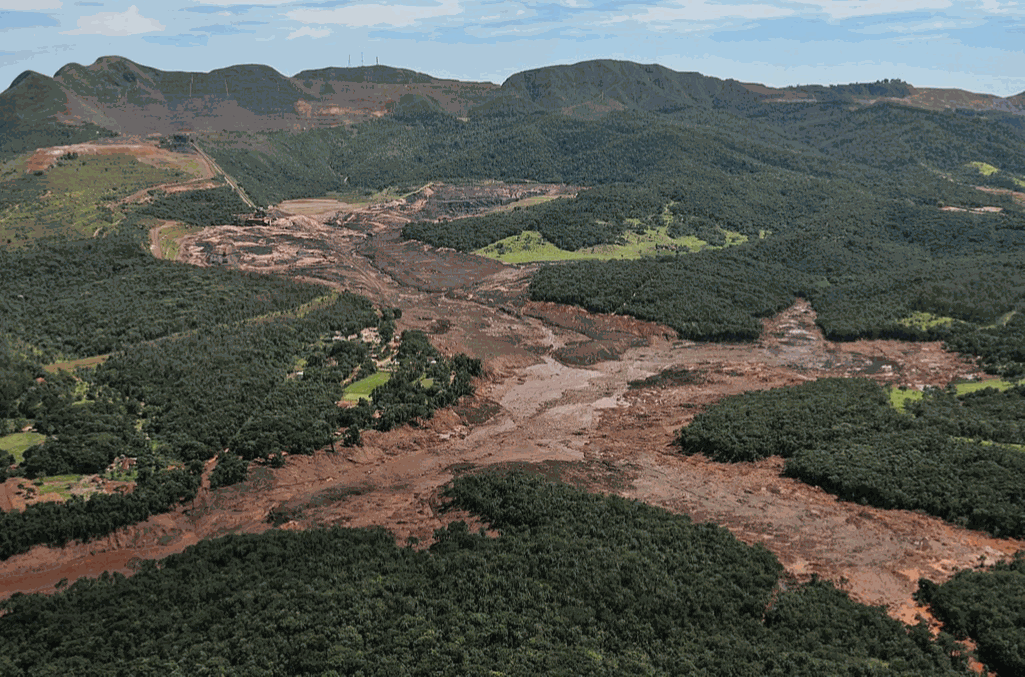}
    \label{fig:brumadinho_real_photo}
}
\hfill
\subfloat[Satellite view of the region and observed flooded area on January 29th, 2019.]{
    \includegraphics[width=0.48\textwidth]{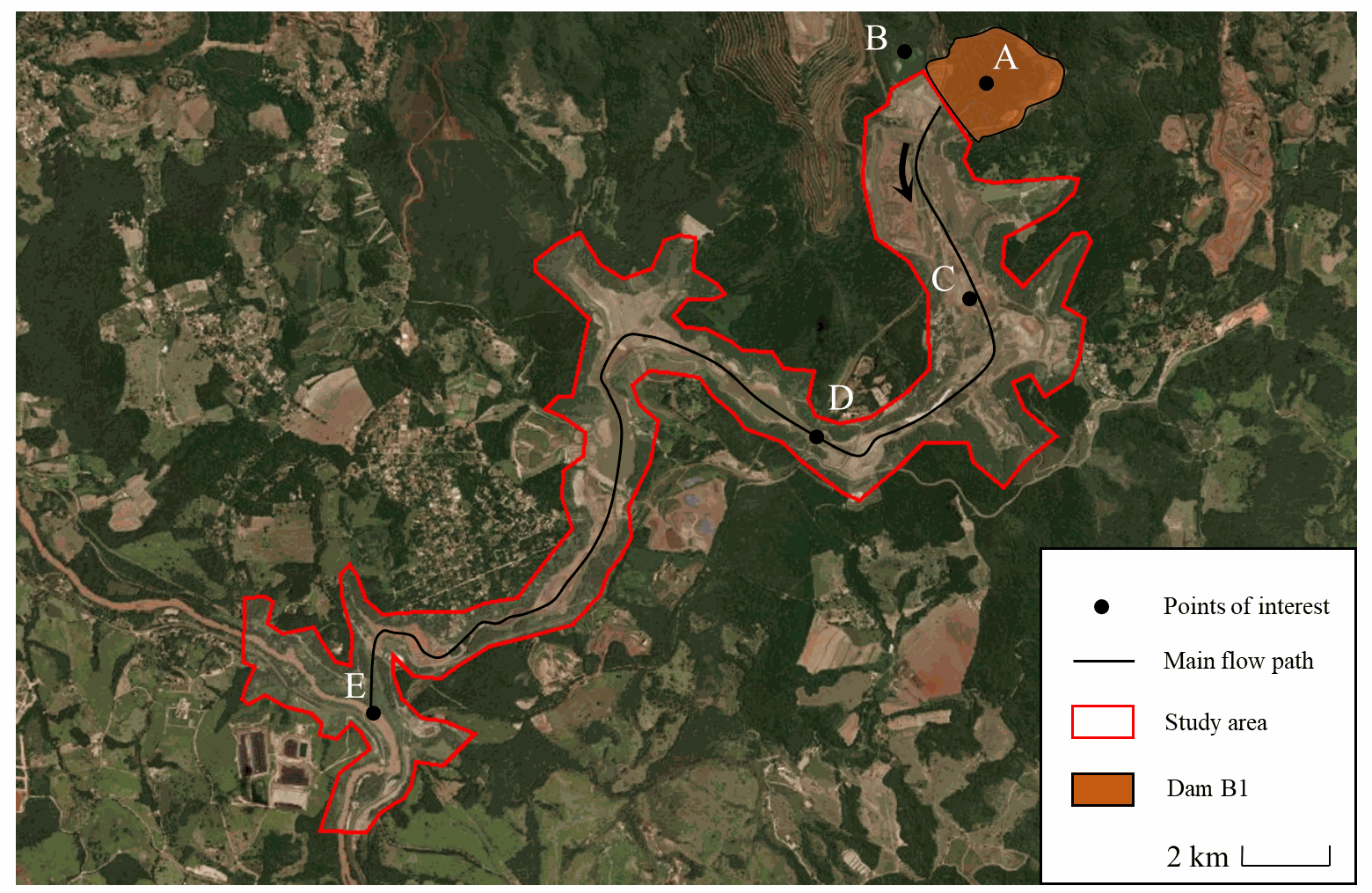}
    \label{fig:brumadinho_study_area}
}

\caption{Images of downstream area affected by the collapse of Dam B1.}
\label{fig:brumadinho_photos}
\end{figure}

According to the official report provided by \cite{Robertson2019}, the collapse was caused by static liquefaction, triggered by a critical combination of ongoing internal strains due to creep and a reduction in shear strength associated with suction loss in the unsaturated zone caused by intense rainfall. Although no signs of structural distress had been identified prior to the failure, Dam B1 lacked sufficient internal drainage, allowing a significant portion of the tailings to remain saturated and thereby favouring undrained flow liquefaction. The rupture occurred in less than 10 seconds and can be observed in video footage available online at \url{https://www.youtube.com/watch?v=L1Feq5-y5J0}. The failure released approximately $9.7 \times 10^6 \ \unit{m}^3$, about $83 \%$ of the total stored volume, of material over a relatively short period of around 5 minutes \citep{Lumbroso2021, Mello2025}, generating a rapidly moving tailings flow similar to a mudflow. Figure \ref{fig:brumadinho_real_photo} shows a picture of the deposited tailings in the immediate downstream area, indicating significant mud depth, lateral spreading and narrowing of the topographic valley to the left of the image.

The resulting tailings flow was both rapid and deep. The available information about the flow combines official findings, observational estimates from video footage, eyewitness reports and values inferred from previous back-analyses. According to the official panel, flow depths as high as $30 \ \unit{m}$ were observed during the propagation of the mud wave, while unofficial estimates based on video footage reported velocities of up to $33 \ \unit{m.s}^{-1}$ over the first $500 \ \unit{m}$ of propagation \citep{Petley2019}. Following the literature \citep{Ligier2020, Lumbroso2021, Gibson2022, Melo2022}, five points of interest (or \textit{control points}) were defined and indicated in Fig. \ref{fig:brumadinho_study_area} along with the study area. Points A and B represent, respectively, the breached Dam B1 and the Dam B4, which remained intact. The other three control points are more relevant in terms of hydrodynamical and risk analyses. Point C represents a canteen near the administrative area of the mine, which was a region immediately downstream the dam with major life losses. Point D is railway bridge destroyed by the tailings flow; and point E refers to the arrival area at the Paraopeba River. The tailings wave took approximately 2 and 10 minutes to reach points C and D, respectively according to observations. Finally, anecdotal evidence indicates that the tailings flow reached the Paraopeba River (point E), located $10 \ \unit{km}$ downstream of Dam B1, between 1.5 and 2.0 hours after the collapse.

Previous studies have provided important estimates of maximum flow depth and arrival time at the selected control points based on deterministic back-analyses, as summarized in Tab. \ref{tab:literature_results}. These estimates serve as pointwise comparative references for the results obtained in the present work, particularly for maximum flow depth and arrival time at control points C, D, and E. Most of these deterministic back-analyses were performed by solving the shallow-water equations (SWE), in which non-Newtonian rheological parameters were incorporated into the basal shear stress. An exception is the work of \cite{Yu2020}, which solved the complete three-dimensional continuity and momentum equations instead of adopting the shallow-water hypothesis. The next subsection details the hydrodynamical deterministic model used in this work to simulate the tailings flow produced by the collapse of Dam B1.

\begin{table}[H]
\centering
\caption{Reported maximum flow depth $\maxFlowDepth$ ($\unit{m}$) and arrival time $\arrivalTime$ ($\unit{min}$) at control points C, D, and E from observations and selected studies. The brackets $[(\cdot)_{min}, (\cdot)_{max}]$ correspond to the interval of minimum and maximum values obtained for a given reference.}
\begin{tabular}{lcccccc}
\hline
& \multicolumn{2}{c}{Point C} & \multicolumn{2}{c}{Point D} & \multicolumn{2}{c}{Point E} \\
\cline{2-3} \cline{4-5} \cline{6-7}
Reference & $\maxFlowDepth$ & $\arrivalTime$ & $\maxFlowDepth$ & $\arrivalTime$ & $\maxFlowDepth$ & $\arrivalTime$ \\
\hline
Observations & (-) & 2.0 & (-) & [10.0, 12.0] & (-) & [90.0, 130.0] \\
Ligier (2020) & (-) & [1.25, 2.25] & (-) & [3.75, 9.50] & (-) & [102.0, 175.0] \\
Yu et al. (2020) & (-) & 1.5 & (-) & 4.3 & (-) & 35.0 \\
Lumbroso et al. (2021) & 18.7 & 1.5 & 16.6 & 9.0 & 11.6 & 86.0 \\
Gibson et al. (2022) & 18.3 & 1.5 & 16.4 & 9.25 & 2.8 & 87.8 \\
Melo and Eleutério (2022) & [12.2, 12.5] & 3.0 & [14.7, 17.3] & 8.0 & [3.0, 18.0] & [46.0, 127.0] \\
\hline
\end{tabular}
\label{tab:literature_results}
\end{table}

\subsection{Deterministic model}
\label{subsec:deterministic_model}

In this work, the deterministic numerical code HEC-RAS v6.6 was employed to simulate the tailings flow produced by the Dam B1 collapse. HEC-RAS is widely used and validated for water and tailings dam-breach analyses, as well as other applications involving free-surface hydraulics \citep{Gibson2022, Melo2023}. In terms of modelling the physical phenomenon of tailings dam-breach flow, two main assumptions are made regarding the tailings flow and the released material. The first assumption considers a shallow flow, where the characteristic flow depth length $\depth$ is significantly smaller than the other length scales, allowing to simplify the governing equations by using depth-averaged properties \citep{Castro2019}. This assumption allows to considerably reduce computational costs while retaining accuracy for large-scale quantities of interest, such as flow depth and velocity fields; this is particularly convenient for the usage of probabilistic approaches. The second assumption concerns the rheological modelling of tailings. The released material is modeled as a single bulk fluid described by a single rheological law, i.e. a one-phase approach. One-phase approaches are widely used to simulate hyperconcentrated flows, mudflows and flows caused by tailings dam-breach event \citep{Obrien1988, Pirulli2017, Gibson2021}. This approach is considered to be accurate and parsimonious enough for estimation of macroscopic properties at large scales, e.g. flow depth and velocity. The next item details the governing equations solved by HEC-RAS v6.6.

\subsubsection{Governing equations}
\label{subsubsec:governing_equations}

HEC-RAS v.6.6 models the propagation of the flow by using the discretized shallow-water equations (SWE) system to model the fluid dynamics (Eqs. \ref{eq:continuity}). Further details can be found in HEC-RAS documentation \citep{Hecras66}. The system is solved within a finite-volume computational mesh, producing spatio-temporal fields of depth and velocity, where other properties are derived from those fields. Effects of porosity, wind, wave forcing and Coriolis effect are neglected.
\begin{equation}
\label{eq:continuity}
\begin{gathered}
\frac{\partial \eta}{\partial t} + \gradientVec \cdot (\depth \velocityVec) = 0 \\
\frac{\partial \velocityVec}{\partial t} + (\velocityVec \cdot \gradientVec)\velocityVec
= -\gravityVec \, \cos^2 \phi \, \gradientVec \eta
+ \frac{1}{\depth}\gradientVec \cdot (\nu_t \depth \gradientVec\velocityVec)
- \frac{\shearStress}{\densityMixture \hydraulicRad} \frac{\cos \psi}{\cos \phi}\frac{\velocityVec}{|\velocityVec|}
\end{gathered}
\end{equation}
where $\eta$ is the flow surface elevation, $\depth$ is the free-surface elevation in relation to the terrain (or depth), $\phi$ is the water surface slope, $\psi$ is the inclination angle of the velocity direction, $\velocityVec = (\velocityX,\velocityY)$ is the velocity vector, $\gravityVec$ is the vector of gravity acceleration, $\nu_t$ is the turbulent viscosity, $\shearStress$ is the total basal shear stress, $\densityMixture$ is the mixture density and $\hydraulicRad$ is the hydraulic radius. The mixture density $\densityMixture$ is a function of the volumetric concentration $\volumetricConc$, as follows:
\begin{equation}
\label{eq:density_mixture}
    \densityMixture = \densityWater + (\densitySolid - \densityWater) \volumetricConc
\end{equation}
where $\densityWater = 998 \ \mathrm{kg.m^{-3}}$ is the water density and $\densitySolid = 2650 \ \mathrm{kg.m^{-3}}$ is the solid-particle density.

The basal shear stress $\shearStress$ can be divided in two components. The first component is related to the roughness of the terrain, which is represented by the Manning coefficient $\manning$. The second component is the viscoplastic stress related to the internal friction of the fluid and directly linked to the fluid rheology; here, we use the Bingham rheological model, under the hypothesis of simple shear. Usually, rheological properties are also a function of $\volumetricConc$ \citep{Obrien1988, Major1992, Boger2013, Shakeel2020}; however, here we model the rheological behavior of tailings by directly imposing Bingham parameters. In this case, $\volumetricConc$ is effectively controlling just the fluid density, as shown in Eq. \ref{eq:density_mixture}. The expression of the basal shear stress is written as:
\begin{equation}
\label{eq:basal_shear_stress}
    \shearStress = \frac{\densityMixture \gravity \manning^2}{\hydraulicRad^{1/3}} |\velocityVec|^2 + \yieldStress + \plasticViscosity \shearRate
\end{equation}

where $\yieldStress$ is the yield stress, $\plasticViscosity$ is the plastic viscosity and $\shearRate$ is the shear rate. Among many closure expressions for the shear rate \citep{Pastor2015, Sao2023}, HEC-RAS v6.6 employs the following expressions as a function of the mean velocity $\bar{V}$ and $\depth$:
\begin{equation}
\label{eq:shear_rate}
\shearRate = \frac{3 \bar{V}}{\depth}
\end{equation}

\subsubsection{Numerical parameters and boundary conditions}
\label{subsubsec:numerical_parameters}

\begin{figure}[H]
    \centering
    \includegraphics[width=0.85\linewidth]{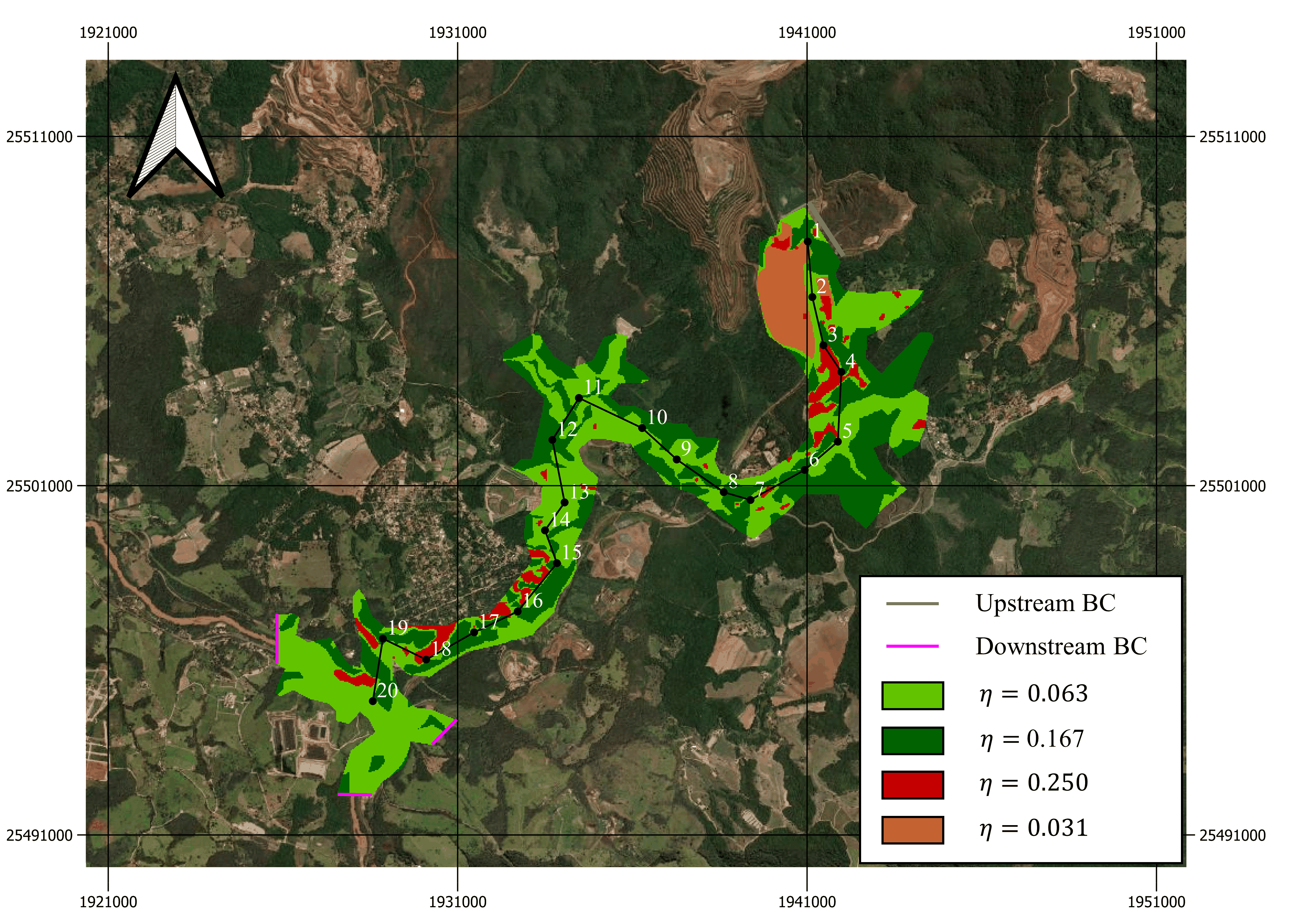}
    \caption{Computational domain, boundary conditions (BCs, in magenta) and areas of different land uses with respective Manning coefficients. Four land uses are defined: forest (dark green, $\manning=0.167$); pasture (light green, $\manning=0.063$), mine workings (brown, $\manning=0.031$) and buildings (red, $\manning=0.250$). Points throughout the main path flow are enumerated from 1 to 20.}
    \label{fig:computational_domain}
\end{figure}

Figure \ref{fig:computational_domain} shows the computational domain covering the study area, boundary conditions and the Manning coefficients associated to the land cover. The domain covers an area of approximately $5.9 \ \unit{km}^2$ and was defined based on the observed flooded area, containing extra area to account for simulations that may exceed the observed flooded area. The mesh was divided in regular cells of size approximately $15 \times 15$ $\unit{m}$, totaling $26212$ cells. For hydrodynamical analyses, the domain was divided into 20 points (P1, ..., P20) along the main flow path, containing control points C (P4), D (P8) and E (P20), as indicated in Fig. \ref{fig:computational_domain}. Four distinct land covers and correspondent Manning coefficients were associated to mesh cells \citep{Lumbroso2021}: forest ($\manning=0.167$); pasture ($\manning=0.063$), mine workings ($\manning=0.031$) and buildings ($\manning=0.250$). The total simulation time was 2 hours (120 minutes) and the time discretization followed the Courant-Friedrichs-Lewy (CFL) condition, with maximum CFL number of 0.8. A maximum of 40 iterations per time step was used.

\begin{figure}[H]
    \centering
    \includegraphics[width=0.7\linewidth]{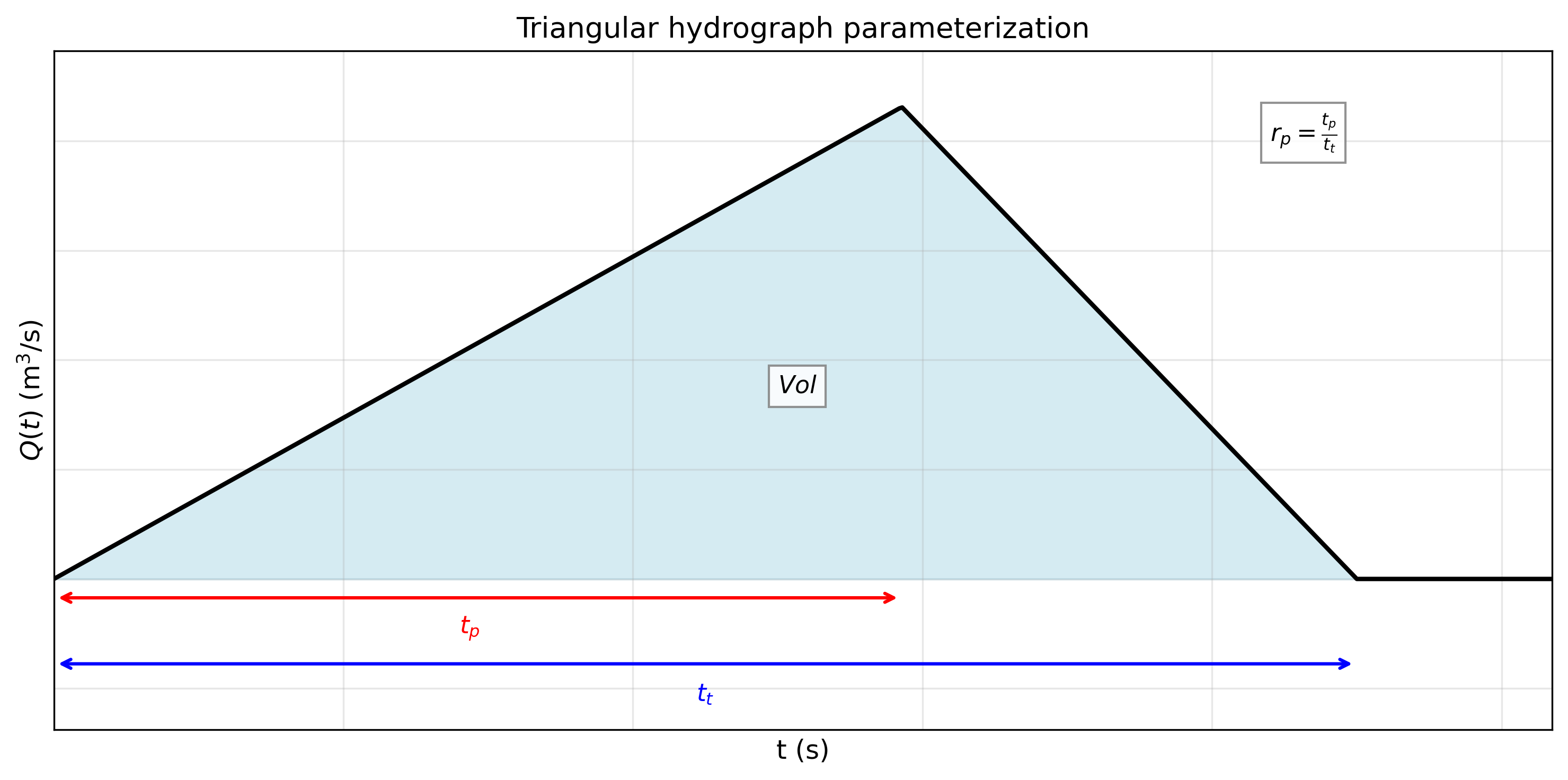}
    \caption{Schematic of triangular hydrograph parameterization.}
    \label{fig:hydrograph}
\end{figure}

Normal-depth conditions were prescribed to downstream boundary conditions, with energy slopes of $0.001 \ \unit{m/m}$, as indicated by Fig. \ref{fig:computational_domain}. Upstream boundary condition was prescribed into the dam's section by defining an inlet hydrograph. In fact, the definition of upstream boundary conditions is inherently challenging because the physical mechanisms involved during the breach formation and volume mobilization are still poorly understood. Experimental efforts were made in this direction \citep{Tian2021, Walsh2021} and a simplified two-fluid model (EMBREA-MUD) was developed by \cite{Petkovvsek2021}. For the Brumadinho dam-breach case, \cite{Lumbroso2021} used EMBREA-MUD to develop a breach hydrograph based on an extreme scenario. The assumptions were: stepwise shape due to the progressive failure; released volume of $9.7 \times 10^6 \ \unit{m}^3$ during 5 minutes; and peak flow rate of $9 \times 10^4 \ \unit{m}^3 \unit{s}^{-1}$. However, this proposed hydrograph reflects adopted modelling assumptions and carries external uncertainties due to EMBREA-MUD modelling, e.g. non-Newtonian rheological parameters, density, erodibility coefficients and geometric parameters of the tailings dam. For the purpose of this work, a triangular hydrograph model (Fig. \ref{fig:hydrograph}) was employed by assuming a parsimonious approximation of the expected phases of the mobilization process, i.e. progressive increase of flow rate until the peak and a subsequent decrease associated with the depletion of mobilizable material. The hydrograph is parameterized in function of three variables: released volume $\releasedVol$; total mobilization time $\totalTimeHyd$ and relative peak time $\relativeTimePeak = t_p / \totalTimeHyd$. The advantage of these variables is that they present easy physical interpretation, while keeping the main physical mechanisms.

\subsubsection{Hazard-related outputs}
\label{subsubsec:outputs}

The present study focuses on a set of hazard-related outputs derived from the simulated transient flow fields, more specifically the flow depth and velocity magnitude fields. Let us denote the flow depth as $\depth_{\indOutputs} (\timeSteps)$ and the depth-averaged velocity fields in a cartesian coordinate system as $\velocityX_{\indOutputs}(\timeSteps)$ and $\velocityY_{\indOutputs}(\timeSteps)$, where $\indOutputs=1,\dots, \nOutputDimensions$ refers to the mesh cell index of the $\nOutputDimensions$-sized computational domain and $n=1,\dots,\numberTimeSteps$ refers to the time index of the $\numberTimeSteps$-sized temporal discretization.

The first output of interest is the temporal development of flow depth at the control points C, D and E. If $\indOutputs_{i}$ denotes the mesh cell associated with control point $i$, the corresponding flow-depth time series is defined as:

$$\depth_{\indOutputs_{i}}(t), \ i \in \{\text{C}, \text{D}, \text{E}\}$$.

They constitute scalar time series and allow the local dynamics of the tailings wave to be analyzed, providing pointwise flow depth that can be compared with available literature back-analyses at the same locations. Flow-depth data were stored at one-minute intervals.

In order to characterize the maximum intensity reached during the event, the maximum flow depth $\maxFlowDepth$ and maximum flow velocity $\maxVel$ are also considered. The quantities $\maxFlowDepth_{\indOutputs}$ and $\maxVel_{\indOutputs}$ are defined at every cell of the computational domain, therefore constituting $\nOutputDimensions$-sized spatial maps. The maximum flow depth and maximum flow velocity are given, respectively, by:
\begin{equation}
    \maxFlowDepth_{\indOutputs} = \max_{1 \leq n \leq \numberTimeSteps} \depth_{\indOutputs} (\timeSteps)
\end{equation}
\begin{equation}
    \maxVel_{\indOutputs} = \max_{1 \leq n \leq \numberTimeSteps} \sqrt{\velocityX_{\indOutputs}(\timeSteps)^2 + \velocityY_{\indOutputs}(\timeSteps)^2}
\end{equation}

The quantity $\maxFlowDepth_{\indOutputs}$ is an indicator of flood severity, since it is directly related to potential damage to people and infrastructure. Likewise, $\maxVel_{\indOutputs}$ provides a measure of flow intensity associated with impact, drag and transport capacity. Data was captured on an interval of 3 seconds to compute $\maxFlowDepth_{\indOutputs}$ and $\maxVel_{\indOutputs}$.

Next, we define the arrival time $\arrivalTime_{\indOutputs}$, since the timing of flooding is critical for hazard assessment. It reflects the available response time and helps distinguish rapidly impacted areas from regions reached only at later stages of the propagation. The arrival time is defined as the first time at which the local flow depth exceeds a prescribed flooding threshold $\depthThresh$:
\begin{equation}
\label{eq:arrival_times}
    \arrivalTime_{\indOutputs} =
\begin{cases}
\min \left\{ \timeSteps : \depth_{\indOutputs}(\timeSteps) \geq \depthThresh \right\},
& \text{if } \exists \ n \text{ such that } \depth_{\indOutputs}(\timeSteps) \geq \depthThresh, \\[4pt]
\text{NaN},
& \text{if } \depth_{\indOutputs}(\timeSteps) < \depthThresh \quad \forall n.
\end{cases}
\end{equation}

Among many possible criteria, we adopt $\depthThresh = 0.31 \ m$, approximately $1 \ \text{ft}$, as suggested by the Federal Emergency Management Agency (FEMA) \citep{FEMA} and HEC-RAS documentation \citep{Hecras66}. If the threshold is never exceeded, the location is considered non-flooded during the simulated event, and $\arrivalTime_{\indOutputs}=\text{NaN}$ is used as a computational convention to identify non-flooded cells. Data was captured at 3-second intervals to compute $\arrivalTime_{\indOutputs}$.

Finally, the flooding probability $P_{f}$ is used to estimate the likelihood of flooding at each cell of the computational domain. This output allows the identification of areas that are systematically flooded, marginally flooded, or flooded only under more unfavorable combinations of input parameters. For a given cell ${\indOutputs}$, the empirical flooding probability is defined as:
\begin{equation}
\label{eq:flooding_probability}
\widehat{P}_{f, \indOutputs}
=
\frac{1}{N}
\sum_{s=1}^{N}
\mathbf{1}\left\{ \maxFlowDepth_{\indOutputs}^{(s)} \geq \depthThresh \right\},
\end{equation}
where $N$ is the number of simulated scenarios, $\maxFlowDepth_{\indOutputs}^{(s)}$ is the maximum flow depth at cell $\indOutputs$ for scenario $s$, and $\mathbf{1}\{\cdot\}$ denotes the indicator function, equal to one if the condition is satisfied and zero otherwise.

\section{Probabilistic framework}

\subsection{Probabilistic modeling of uncertain inputs}
\label{subsubsec:probabilistic_modeling}

By using the deterministic model described in Subsection \ref{subsec:deterministic_model}, the following input variables are modelled as uncertain: volumetric concentration $\volumetricConc$, yield stress $\yieldStress$, plastic viscosity $\plasticViscosity$, relative time of peak $\relativeTimePeak$, total hydrograph time $\totalTimeHyd$ and released volume $\releasedVol$. Uniform distributions were attributed independently to the input variables to model their uncertainty, since we assume the same probability of them occurring due to lack of knowledge and, at the same time, avoid introducing unknown information to the distribution. The uniform distributions are constrained to a range of minimum and maximum values in order to represent physically the specific case of Brumadinho dam-breach. In the specific context of physical properties, there is a lack of systematic data in literature containing rheological properties of tailings in a post-failure condition and following the same rheological measurement protocol. In this case, calibrated values explored by \cite{Ligier2020} and \cite{Lumbroso2021} were used as maximum and minimum reference values. For hydrograph parameters, uncertainty introduced to $\totalTimeHyd$ and $\relativeTimePeak$ was based on information of \cite{Lumbroso2021} and \cite{Mello2025}, whereas $\releasedVol$ uncertainty is based on previous studies related to the Brumadinho disaster \citep{Robertson2019, Mello2025}.

\begin{table}[h!]
\centering
\caption{Probabilistic modeling of uncertain input variables and justification of statistical supports.}
\label{tab:PDFs}
\small
\begin{tabularx}{\textwidth}{l l l X}
\toprule
\textbf{Input variable} & \textbf{PDF and support} & \textbf{Unit} & \textbf{Justification of support} \\
\midrule
$\volumetricConc$  & $\mathcal{U}\!\left[20,\, 60 \right]$ & $\unit{\%}$ 
& From \citep{Obrien1988, Major1992, Fitton2012}. \\

$\yieldStress$      & $\mathcal{U}\!\left[250,\, 1725\right]$ & $\unit{Pa}$ 
& From \citep{Ligier2020, Lumbroso2021}. \\

$\plasticViscosity$       & $\mathcal{U}\!\left[30,\, 100\right]$  & $\unit{Pa \ s}$
& From \citep{Ligier2020, Lumbroso2021}. \\

$\relativeTimePeak$        & $\mathcal{U}\!\left[0.05,\, 0.95\right]$  & $\unit{-}$ 
& Constrained dimensionless quantity. \\

$\totalTimeHyd$       & $\mathcal{U}\!\left[200,\, 400\right]$ & $\unit{s}$ 
& From \citep{Robertson2019, Lumbroso2021, Mello2025}. \\

$\releasedVol$      & $\mathcal{U}\!\left[9.74,\, 9.90\right] \times 10^6$  & $\unit{m}^3$ 
& From \citep{Robertson2019, Lumbroso2021, Mello2025}. \\

\bottomrule
\end{tabularx}
\end{table}

\subsection{Metamodeling for high-dimensional outputs}
\label{subsubsec:PCA_GPR}

In order to predict sufficient hazard-related output maps to perform a probabilistic analysis, two main issues must be addressed: the high-dimensionality of output maps; and the relatively high computational cost of the deterministic code. The first issue is solved by using a dimensionality-reduction technique, which consists of expanding the high-dimensional outputs into a reduced linear basis. Therefore, data is projected to a reduced-order space and represented as basis coefficients. The second issue is addressed by training a fast-to-evaluate statistical model, i.e a \textit{metamodel} or surrogate model, with few deterministic points. The main logic of this framework is to train metamodels to predict basis coefficients. These predicted basis coefficients can be reprojected to the original high-dimensional space and, ultimately, form the predicted maps. Several works explored this joint application of dimensionality-reduction and metamodeling \citep{Li2020, Elgarroussi2022, Rohmer2023}. Here, we employ principal component analysis (PCA) and Gaussian process regression (GPR) as dimensionality-reduction and metamodeling techniques, respectively. A brief description of both techniques is given in the following items.

\subsubsection{Sampling the training and validation sets}
\label{subsubsec:training_set}

The probabilistic modeling of inputs presented in subsection \ref{subsubsec:probabilistic_modeling} allows to sample scenarios from the joint PDF to define a training and validation set. Latin-hypercube sampling (LHS) technique is used to sample the sets, due to its capability to fill the input space adequately, while avoiding clustering and redundancy of input data \citep{Pronzato2012}. Here, $550$ scenarios were sampled and divided into training and validation sets. The training set contains scenarios used to train metamodels, whereas the validation set is employed to evaluate the predictive performance of the metamodels. Different training sizes were tested with a random holdout validation with 10 repetitions, ranging from $150$ to $450$ training scenarios at steps of $50$ scenarios. The testing results are shown in the Supplementary Material; the complete set of $450$ simulations was retained to maximize predictive performance. Selection bias of the training set is negligible according to the testing results. Therefore, a split of $450/100$ training/validation scenarios was used for the probabilistic framework. Each scenario was evaluated by the HEC-RAS deterministic model, producing high-dimensional maps as outputs (see Subsection \ref{subsec:deterministic_model}). The next subsection addresses the dimensionality-reduction of training outputs with PCA technique.

\subsubsection{Principal component analysis}
\label{subsubsec:PCA}

\newcommand{\dataMatrixPCA}{\mathcal{Y}}
\newcommand{\centeredData}{\widetilde{\dataMatrixPCA}}
\newcommand{\covarianceMatrixPCA}{\Sigma}
\newcommand{\coefficientsMatrixPCA}{C}
\newcommand{\basisMatrixPCA}{V}
\newcommand{\dimensionMean}{\overline{\outputPoint}}
\newcommand{\outputPoint}{y}

Principal component analysis (PCA) is a data-driven dimensionality-reduction technique, which provides a linear representation of high-dimensional data in a reduced space. For more details, we refer to \cite{Abdi2010}. Let $\dataMatrixPCA = \left[ \outputPoint^{(1)}, \dots, \outputPoint^{(\nDoE)} \right]^{\top} \in \R^{\nDoE \times \nOutputDimensions}$, where $\outputPoint^{(\indPoints)} = \outputPoint(\inputPoint^{(\indPoints)})$ is the $\indPoints$-th high-dimensional output obtained by the deterministic simulator. The dimension-wise mean $\dimensionMean \in \R^{\nOutputDimensions}$ is calculated as:
\begin{equation}
\label{Eq:dimension_wise_mean}
    \dimensionMean = \frac{1}{\nDoE} \sum^{\nDoE}_{\indPoints=1} \outputPoint^{(\indPoints)}
\end{equation}

Centered data $\centeredData \in \R^{\nDoE \times \nOutputDimensions}$ is obtained by subtracting $\overline{\outputPoint}$ from each row of $\dataMatrixPCA$. Then, the empirical covariance matrix $\covarianceMatrixPCA \in \R^{\nOutputDimensions \times \nOutputDimensions}$ is calculated based on $\centeredData$:
\begin{equation}
\label{Eq:covariance_matrix_PCA}
    \covarianceMatrixPCA = \frac{1}{\nDoE - 1} \centeredData^{\top} \centeredData
\end{equation}

The eigenvalue problem $\covarianceMatrixPCA \basis_{\indBasis} = \lambda_{\indBasis} \basis_{\indBasis}$ is solved and the eigenvalues are sorted in decreasing order $\lambda_1 \geq ... \geq \lambda_{\nOutputDimensions}$, representing the data variance explained by the corresponding eigenvectors. The eigenvector associated with the $\indBasis$-th eigenvalue is denoted by $\basis_{\indBasis} = (\basis_{\indBasis,1}, \dots, \basis_{\indBasis, \nOutputDimensions})^{\top} \in \R^{\nOutputDimensions}$. The truncated PCA basis is obtained by retaining the first $\nbasis$ eigenvectors, with $\nbasis \ll \nOutputDimensions$, and is denoted by $\basisMatrixPCA = (\basis_{\indBasis, \indOutputs})_{1 \leq \indBasis \leq \nbasis, \ 1 \leq \indOutputs \leq \nOutputDimensions} \in \R^{\nbasis \times \nOutputDimensions}$, where the rows of $\basisMatrixPCA$ contain the retained basis vectors. The Gram matrix can be defined as $\GramMatrix = \basisMatrixPCA \basisMatrixPCA^{\top}$. The PCA basis coefficients are obtained by projecting the centered data onto the truncated basis $\coefficientsMatrixPCA = \centeredData \basisMatrixPCA^{\top} \in \R^{\nDoE \times \nbasis}$. Equivalently, $\coefficientsMatrixPCA = ( \coef_{\indBasis}(\inputPoint^{(\indPoints)}) )_{1 \leq \indPoints \leq \nDoE, \ 1 \leq \indBasis \leq \nbasis}$, where the $\indPoints$-th row of $\coefficientsMatrixPCA$ contains the basis coefficients associated with the input scenario $\inputPoint^{(\indPoints)}$: $\coef(\inputPoint^{(\indPoints)}) = (\coef_1(\inputPoint^{(\indPoints)}),\ldots,\coef_{\nbasis}(\inputPoint^{(\indPoints)}))^\top \in \R^{\nbasis}$. Finally, the $\indOutputs$-th output dimension of the $\indPoints$-th original map can be represented as:
\begin{equation}
\label{Eq:PCA_reconstruction}
    \outputPoint_{\indOutputs}^{(i)} \approx \dimensionMean_{\indOutputs} + \sum^{\nbasis}_{\indBasis=1} \coef_{\indBasis} (\inputPoint^{(i)}) \basis_{\indBasis,\indOutputs}
\end{equation}

All output dimensions for the $\indPoints$-th map can be computed directly in vector form as follows: 
\begin{equation}
\label{Eq:PCA_reconstruction_map}
    \outputPoint^{(\indPoints)} \approx \dimensionMean + \basisMatrixPCA^{\top} \coef(\inputPoint^{(\indPoints)})
\end{equation}

The PCA procedure was implemented in Python using the \textit{sklearn.decomposition} module of \textit{scikit-learn} library. Here, each hazard-related output required a different number of retained principal components: for maximum flow depth, $\nbasis = 10$ components ($99\%$ explained variance); for maximum flow velocity, $\nbasis = 5$ components ($99\%$ explained variance); and for arrival times, $\nbasis = 60$ components ($96\%$ explained variance). Screeplots and estimates of PCA truncation error are available in the Supplementary Material.

We note that standard PCA requires a complete data matrix for the aforementioned operations and principal components cannot be defined in the presence of missing values. This problem is observed specifically for dimensionality reduction of arrival times maps (see Eq. \ref{eq:arrival_times}), where the data matrix presents undefined values at non-flooded cells. In this work, this is solved by attributing the expected value of a given cell to its correspondent missing values. This practice is often called \textit{mean-imputation}. Even though it does not introduce new information for each individual point, since PCA is performed on centered variables, it underestimates the total variance in the dataset \citep{Gomez2014}. Specifically in this work, the mean-imputation technique adds non-physical values to the dataset, i.e. arrival times at non-flooded cells. Therefore, in any case, those points should be removed afterwards, since they carry no physical information; a classification technique is presented in item \ref{subsubsec:RFC} for this end. 

\subsubsection{Gaussian process regression}
\label{subsubsec:GPR}

\newcommand{\meanGPR}{m}
\newcommand{\kernelGPR}{K}
\newcommand{\DoePoint}{\mathrm{x}}

\renewcommand{\indGPR}{j}
\renewcommand{\GaussianProcess}{Z} 
\renewcommand{\MeanFunction}{m}
\renewcommand{\KernelFunction}{K}
\renewcommand{\condGP}{\GaussianProcess_n}
\renewcommand{\ConditionalMean}{\MeanFunction_c}
\renewcommand{\ConditionalKernel}{\KernelFunction_c}
\renewcommand{\condGPVector}{\GaussianProcess_n}
\renewcommand{\ObservationVector}{y^{\mathrm{obs}}}
\renewcommand{\KernelMatrix}{\KernelFunction(\DoE, \DoE)}
\renewcommand{\KernelVector}[1]{\KernelFunction(\DoE, #1)}
\renewcommand{\KernelVectorRow}[1]{\KernelFunction(#1, \DoE)}
\renewcommand{\NumberGPite}{N_{Z}}
\renewcommand{\nX}{N_X}
\newcommand{\deterministicModelGPR}{g}

Gaussian process regression (GPR) is a probabilistic metamodeling technique used to approximate the relationship between an input vector and a scalar model response. For more details, we refer to \cite{Williams1995}; some examples of applications that use GPR can be found in \citep{Cui2021, Zhu2025}. Let $\deterministicModelGPR : \ \mathbb{X} \subseteq \R^\nvar \longrightarrow \R$ be a multivariate function, modeling a deterministic computer code. As a prior assumption, assume that $\deterministicModelGPR$ is one trajectory of a Gaussian process $Z$ on $\mathbb{X}$, denoted as $Z \sim \mathcal{GP}(\meanGPR,\kernelGPR)$, where $\meanGPR$ is the mean function and $\kernelGPR$ is the covariance function, or kernel. This prior information can be updated using a DoE $\DoE$ and respective observations $\ObservationVector_{\indPoints} = \deterministicModelGPR(\inputPoint^{(\indPoints)})$. This results in a posterior random process, that remains Gaussian. That conditional GP is denoted $\condGP \sim \mathcal{GP}(\meanGPR_c, \kernelGPR_c)$, where $\meanGPR_c$ and $\kernelGPR_c$ are given in closed-form by:
\begin{equation}
\begin{aligned}
\ConditionalMean(\inputPoint) &= \meanGPR(\inputPoint) + \KernelVectorRow{\inputPoint} \KernelMatrix^{-1} (\ObservationVector - \meanGPR(\DoE))  \qquad (\inputPoint \in \mathbb{X}) \\
\ConditionalKernel(\inputPoint, \inputPoint') &= \KernelFunction(\inputPoint, \inputPoint') - \KernelVectorRow{\inputPoint} \KernelFunction(\DoE, \DoE)^{-1} \KernelVector{\inputPoint'} \qquad (\inputPoint, \inputPoint' \in \mathbb{X})
\end{aligned}
\label{eq:kriging_mean_variance_intro}
\end{equation}
where $\ObservationVector$ is the vector of observation points, $\KernelMatrix = (\KernelFunction(\inputPoint^{(\indPoints)}, \inputPoint^{(j)}))_{1 \leq \indPoints,j \leq \nDoE}$ is the covariance matrix at design points, $\KernelVector{\inputPoint} = (\KernelFunction(\inputPoint^{(\indPoints)}, \inputPoint))_{1 \leq \indPoints \leq \nDoE}$ is the column vector of covariances between a new point $x$ and the design points, and $\KernelVectorRow{\inputPoint} = \KernelVector{\inputPoint}^\top$.

Here, the scalar response modeled by GPR corresponds to a PCA basis coefficient, i.e $\deterministicModelGPR_\indBasis(\inputPoint) = \coef_{\indBasis}(\inputPoint)$ for each $\indBasis$-th retained basis component. The associated observation vector is the $\indBasis$-th column of the coefficient matrix: $\coefficientsMatrixPCA_{.,\indBasis} = (\coef_{\indBasis}(\inputPoint^{(\indPoints)}))_{\indPoints=1,\dots,\nDoE} \in \R^{\nDoE}$. One scalar GPR metamodel is then trained for each coefficient function $\inputPoint \mapsto \coef_\indBasis(\inputPoint)$, using the DoE $\DoE$ and the observations $\coefficientsMatrixPCA_{\cdot,\indBasis}$ as training data. Thus, a set of $\nbasis$ independent GPR metamodels is constructed: $Z_n^{(\indBasis)} \sim \mathcal{GP}(\ConditionalMean^{(\indBasis)}, \ConditionalKernel^{(\indBasis)})$, with $\indBasis=1,\dots, \nbasis$. In the Appendix, Algorithm \ref{algo:prediction_intro} presents the PCA–GPR procedure, divided into an offline training phase and an online prediction phase. Both PCA and GPR procedures introduce errors which may propagate to the reconstructed maps; their explicit propagation is outside the scope of this study and remains a limitation of the present analysis.

Independent GPR metamodels were constructed for each retained PCA coefficient using the Python library \textit{GPflow} \citep{Matthews2017}. A zero mean function and a Matérn 5/2 covariance kernel with one length scale per input variable were adopted. For each metamodel, the kernel variance was initialized using the empirical variance of the corresponding PCA coefficient, while the length scales were initialized using the mean values of the training input sample. The kernel variance, length scales and Gaussian-likelihood noise variance were then estimated by maximizing the marginal likelihood using the \textit{GPflow SciPy} optimizer. Validation of metamodels is carried out by evaluating predictive performance metrics, such as the Nash-Sutcliffe efficiency ($Q^2$) and the root mean squared error ($\mathrm{RMSE}$) \citep{Marrel2011, Perrin2021}. They are defined and presented for all hazard-related properties in the Supplementary Material; some representative results are shown in Appendix \ref{app:validation}. Both metrics were submitted to a repeated random holdout validation, with 10 repetitions, in order to assess variability with respect to the random choice of training/validation scenarios. In each repetition the available deterministic scenarios were randomly partitioned into $450$ training scenarios and $100$ independent validation scenarios; then, the correspondent median and standard deviation are reported.

\subsubsection{Random forest classification for data-filtering}
\label{subsubsec:RFC}

Some hazard-related outputs of this work, more specifically the pointwise flow depth ($\depth_{\indOutputs_{i}}(t), \ i \in \{\text{C}, \text{D}, \text{E}\}$) and arrival times maps ($\arrivalTime_{\indOutputs}$) require specific data treatment for physically-consistent predictions, because they are affected by temporal discontinuities of flooded/non-flooded states. Thus, filtering out non-flooded states is essential to obtain more accurate predictions and to retain physical consistency. More details on how flooded/non-flooded states affect each hazard-related output are given in subsection \ref{subsec:treatment_outputs}. Here, an auxiliary Random Forest Classifier (RFC) is employed to classify flooded and non-flooded states according to input scenarios. RFCs are interesting due to their efficient performance in high-dimensional spaces and has a relatively low computational cost \citep{Feng2015, Farhadi2021}. An RFC is an ensemble learning method composed of multiple decision trees, each trained from bootstrap samples of the training data and random subsets of predictors. For a given input scenario, the final class is assigned by majority voting among the trees, ultimately providing the most likely results. More details can be found in \cite{Breiman2001} and \cite{Salman2024}. Here, the RFC procedure was implemented in Python using the \textit{RandomForestClassifier} from the \textit{sklearn.ensemble} module; validation metrics for RFC are defined and shown in the Supplementary Material.

\subsection{Global sensitivity analysis}
\label{subsec:GSA}

Variance-based global sensitivity analysis (GSA) is employed to quantify how uncertainties in the model inputs contribute to the variability of the predicted hazard outputs. Let $\inputVectorRandom=(\inputPointRandom{1},\ldots,\inputPointRandom{\nvar})\in\mathbb{R}^{\nvar}$ denote the vector of independent random inputs and let $Y=\deterministicModel(\inputVectorRandom)$ be a scalar model response. The variance of $Y$ may be represented as $\totalVariance = \Var(Y)$. For a subset of inputs $\indexSet\subseteq\{1,\ldots,\nvar\}$, the unnormalized closed and total variance contributions (unnormalized Sobol' indices) are defined, respectively, as
\begin{equation}
\label{eq:sobol_closed_total}
D_{\indexSet}^{\mathrm{C}}
=
\Var\left(
\Esp\left[
Y\mid\inputVectorRandom_{\indexSet}
\right]
\right),
\qquad
D_{\indexSet}^{\mathrm{T}}
=
\totalVariance
-
\Var\left(
\Esp\left[
Y\mid\inputVectorRandom_{\exceptIndexSet}
\right]
\right),
\end{equation}
where $\exceptIndexSet$ denotes the complementary set of inputs. Their normalized forms (normalized Sobol' indices) are
\begin{equation}
\label{eq:sobol_normalized}
\sobolClosed{\indexSet}
=
\frac{D_{\indexSet}^{\mathrm{C}}}{\Var(Y)},
\qquad
\sobolTot{\indexSet}
=
\frac{D_{\indexSet}^{\mathrm{T}}}{\Var(Y)}.
\end{equation}
For a single input $\indexSet=\{i\}$, $\sobolClosed{i}$ corresponds to its first-order effect (or main effect), whereas $\sobolTot{i}$ accounts for its main effect and all its interactions with the remaining inputs. Further details on variance-based GSA and Sobol' indices can be found in \cite{Saltelli2008, DaVeiga2021}.

In practice, the indices are generally estimated using pick-freeze (PF) schemes based on two independent samples of the input vector. To estimate the influence of a selected input or group of inputs, the corresponding variables are frozen from one sample while the complementary variables are taken from the other sample \citep{Saltelli2008}. The paired model evaluations are then used to estimate the closed and total variance contributions in Eq.~\ref{eq:sobol_closed_total} by using empirical estimators. For first-order and total indices, we employ the Janon-Monod estimator \citep{Janon2014} and the Jansen estimator \citep{Jansen1999}, respectively. In the present framework, these evaluations are obtained from the trained GPR metamodels \citep{Janon2014_2}, allowing sufficiently large PF samples to be generated without additional HEC-RAS simulations.

For a high-dimensional response $\mathbf{Y}(\inputVectorRandom) =(Y_1(\inputVectorRandom),\ldots,Y_{\nOutputDimensions}(\inputVectorRandom))^\top$, with $\indOutputs = 1, \dots, \nOutputDimensions$, the PF estimators can first be applied independently to each output dimension. This strategy is referred to here as the \textit{dimension-wise approach}. For the output dimension $\ell$, the closed and total indices are estimated with their corresponding PF estimator, denoted as $\pfEstim{(\cdot)}$ as
\begin{equation}
\label{eq:dimensionwise_indices}
\pfEstim{\sobolClosed{\indexSet, \indOutputs}}
=
\frac{\pfEstim{\SobolUnnormalized{\indexSet, \indOutputs}^{c}}}
{\pfEstim{\totalVariance_{\indOutputs}}},
\qquad
\pfEstim{\sobolTot{\indexSet, \indOutputs}}
=
\frac{\pfEstim{\sobolTotD{\indexSet, \indOutputs}}}
{\pfEstim{\totalVariance_{\indOutputs}}}.
\end{equation}
Applying Eq.~\ref{eq:dimensionwise_indices} over all mesh cells produces sensitivity maps, whereas its application over successive time instants produces time-dependent sensitivity curves \citep{Marrel2011}. This approach is particularly appropriate when the final predicted response includes dimension-specific (e.g. the scalar response of a single output dimension) or nonlinear treatments, such as flooded/non-flooded classification, which cannot be represented solely through a linear PCA reconstruction.

When the high-dimensional response is fully represented by the retained PCA expansion, a \textit{basis-derived approach} can instead be employed. Consider a PCA reconstruction of the GPR-predicted response, similar to Eq. \ref{Eq:PCA_reconstruction}:
\begin{equation}
\label{Eq:PCA_reconstruction_GSA}
    \outputPoint_{\indOutputs}^{*(i)} \approx \dimensionMean_{\indOutputs} + \sum^{\nbasis}_{\indBasis=1} \coef^{*}_{\indBasis} (\inputPoint^{(i)}) \basis_{\indBasis,\indOutputs}
\end{equation}

where $(\cdot)^{*}$ represents a GPR-predicted value.

Following \citep{Li2020, Sao2025}, the basis-derived closed-form formula for the Sobol' index of the $\indOutputs$-th output dimension $\pfEstim{\sobolClosed{\indexSet, \indOutputs}} \in \R$ is written as:
\begin{equation}
\label{eq:GSA_reprojection}
    \pfEstim{\sobolClosed{\indexSet, \indOutputs}} = \frac{\basisVec^{\top} \ \pfEstim{\sobolClosedD{\indexSet}}(\coefPred) \ \basisVec}{\basisVec^{\top} \ \pfEstim{\totalVariance}(\coefPred) \ \basisVec}
\end{equation}

where $\pfEstim{\sobolClosedD{\indexSet}}(\coefPred) \in \R^{\nbasis \times \nbasis}$ is the matrix of unnormalized Sobol' indices of the vector of predicted basis coefficients and $\pfEstim{\totalVariance}(\coefPred) \in \R^{\nbasis \times \nbasis}$ is the covariance matrix of the vector of predicted basis coefficients. Total indices are estimated similarly, by replacing $\pfEstim{\sobolClosedD{\indexSet}}(\coefPred)$ by $\pfEstim{\sobolTotD{\indexSet}}(\coefPred)$, as shown in \citep{Sao2025}. A vector-valued PF scheme should be used to calculate those matrices and further details are described in \citep{Sao2025}.

Since PF estimation is performed in a reduced space of dimension $\nbasis \ll \nOutputDimensions$, this approach considerably reduces the computational and memory requirements associated with high-dimensional outputs \citep{Li2020,Sao2025}. The basis-derived approach is therefore adopted when the retained PCA expansion fully represents the analyzed output, whereas the dimension-wise approach is used when the final prediction involves nonlinear or output-specific treatment. The prediction and GSA strategies selected for each output are summarized in Subsection~\ref{subsec:treatment_outputs}.

\subsection{Summary of implementation of the probabilistic framework for each output}
\label{subsec:treatment_outputs}

Each output of interest requires a specific treatment because of its data structure and flooded/non-flooded behavior. Table \ref{tab:uq_gsa_workflows} presents a summary of the methods used for prediction and to perform GSA for the different outputs. Validation of the employed methods can be found in Appendix \ref{app:validation}. Both $\arrivalTime_{\indOutputs}$ and $\depth_{\indOutputs_{i}}(t), \ i \in \{\text{C}, \text{D}, \text{E}\}$ require further treatment before the GSA procedure. 

In the case of $\arrivalTime_{\indOutputs}$, mean-imputation is used as pre-treatment to substitute NaN values for the cellwise expected value calculated from the flooded training scenarios, allowing the application of PCA and GPR. Although mean imputation enables the application of PCA, it removes the information distinguishing flooded from non-flooded cells and assigns artificial arrival-time values to non-flooded cells. This information is recovered using cellwise RFCs trained for cells with scenario-dependent flood status. For each predicted scenario, the classifiers identify the non-flooded cells and generate a mask that is applied to the PCA--GPR arrival-time prediction. Then, the set of RFCs produces a mask of non-flooded cells for each prediction, which is used for the post-treatment of predicted $\arrivalTime_{\indOutputs}$. The masking guarantees that the final map presents physically-consistent values only. For GSA, the dimension-wise Sobol' indices of arrival time are computed only at cells classified as flooded in all model evaluations entering the corresponding PF estimator, including the two independent samples and the hybrid samples. Indices outside this common flooded support are left undefined and masked. Consequently, the resulting sensitivity maps characterize variations in arrival time without mixing them with variations in flood occurrence.

For $\depth_{\indOutputs_{i}}(t), \ i \in \{\text{C}, \text{D}, \text{E}\}$, control points are associated to time-series. At each time step, the response has a mixed structure comprising a point mass at zero for non-flooded scenarios and a continuous distribution of positive flow depths for flooded scenarios. Such a response is not adequately represented by a single GPR with a standard stationary kernel. Therefore, an RFC first predicts the flooded state, and the time-wise GPR predicts the flow depth only for scenarios classified as flooded.

\begin{table}[htbp]
\centering
\caption{Summary of the output-specific joint UQ-GSA workflows adopted in this work.}
\label{tab:uq_gsa_workflows}
\renewcommand{\arraystretch}{1.2}
\begin{tabularx}{\textwidth}{>{\raggedright\arraybackslash}p{3.2cm}
                                >{\raggedright\arraybackslash}X
                                >{\raggedright\arraybackslash}X
                                >{\raggedright\arraybackslash}X
                                >{\raggedright\arraybackslash}p{3.2cm}}
\toprule
\textbf{Output} & \textbf{Pre-treatment} & \textbf{Prediction} & \textbf{Post-treatment} & \textbf{GSA technique} \\
\midrule
$\maxFlowDepth_{\indOutputs}$
& (-)
& PCA + GPR 
& (-) 
& Basis-derived approach \\

$\maxVel_{\indOutputs}$ 
& (-)
& PCA + GPR 
& (-) 
& Basis-derived approach \\

$\arrivalTime_{\indOutputs}$ 
& Mean-imputation
& PCA + GPR 
& RFC masking of non-flooded cells 
& Dimension-wise approach \\

$\depth_{\indOutputs_{i}}(t), \ i \in \{\text{C}, \text{D}, \text{E}\}$ 
& Removal of non-flooded scenarios for GPR training
& RFC + Time-wise GPR 
& (-) 
& Dimension-wise approach \\
\bottomrule
\end{tabularx}
\end{table}

\section{Results and discussion}

\subsection{Probabilistic flooded area}
\label{subsec:probabilistic_flooded_area}

\begin{figure}[H]
    \centering
    \includegraphics[width=0.9\linewidth]{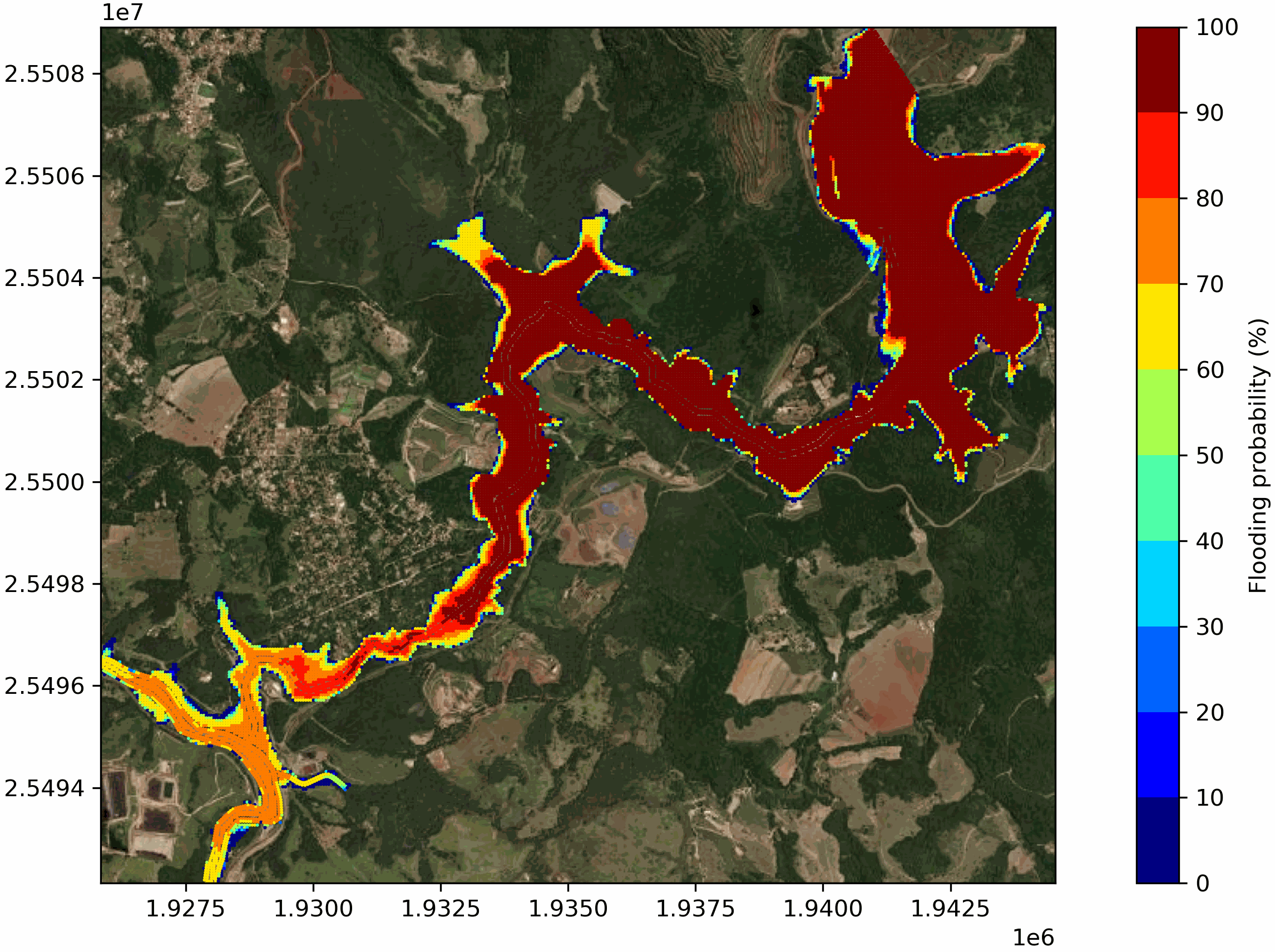}
    \caption{Map of flooding probability calculated from $5000$ scenarios. Values close to $100 \%$ (hot colors) represent areas that are flooded in almost all scenarios; values close to $0 \ \%$ (cold colors) indicate areas that are rarely reached by the flow.}
    \label{fig:flooding_probability}
\end{figure}

Figure \ref{fig:flooding_probability} shows the probabilistic flooded map, which is useful to identify where the flow is expected to reach with potential to create risk. The map is calculated with Eq. \ref{eq:flooding_probability}, where $5000$ predicted scenarios of maximum flow depth were used and cells with $\maxFlowDepth_{\indOutputs} > \depthThresh$ were considered as flooded. No significant differences were observed by calculating the probabilistic maps by filtering the maximum flow depth maps and by using RFCs at each cell. In general, the map presents a significant area of high probability of flooding (between $90 - 100\%$), encompassing the Córrego do Feijão mine administrative area and subsequent flow path within the relatively narrow topographic valley. It means that these regions were very susceptible to flooding, independently of the explored scenario within the parameters' range defined by Tab. \ref{tab:PDFs}. Downstream regions show intermediate flooding probability, indicating stronger uncertainty in how far and how widely the tailings flow propagates. These results support the usage of probabilistic approaches. A deterministic approach would omit areas that are conditionally affected (intermediate and low flooding probability); furthermore, from a risk perspective, a probabilistic approach identify areas where uncertainty should be explicitly considered in hazard assessment. 

\subsection{Spatial development of outputs along the flow path}
\label{subsec:spatial_development}

\begin{figure}[H]
    \centering
    \includegraphics[width=0.9\linewidth]{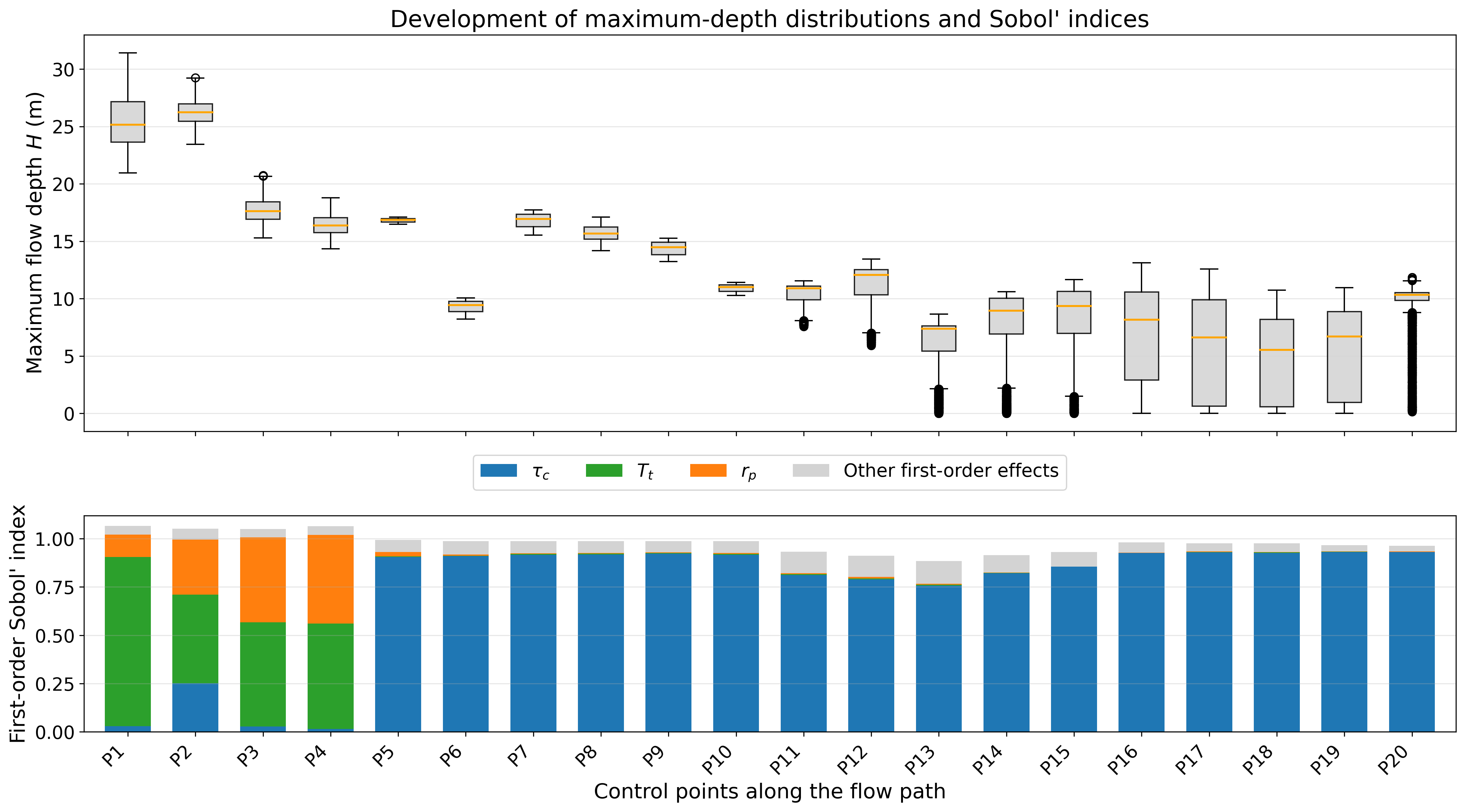}
    \caption{Spatial development of $\maxFlowDepth$ over 20 points along the main flow path. Distributions of data and correspondent Sobol' indices are shown in the top and bottom panels, respectively. $10000$ input scenarios were used to generate the output distributions and PF samples of size $10000$ were used for Sobol' indices estimation.}
    \label{fig:spatial_development_H}
\end{figure}

The spatial development of hazard-related properties can be evaluated along the main flow path, discretized in 20 points defined in Fig. \ref{fig:computational_domain}. For a given property, each point is associated with a distribution represented by a boxplot. For $\maxFlowDepth$ and $\maxVel$, $10000$ predicted scenarios were used and only flooded scenarios were included in the distributions. For $\arrivalTime$, $5000$ predicted scenarios were used. Then, variability of each point is explained by the Sobol' indices. First, Fig. \ref{fig:spatial_development_H} shows the results for $\maxFlowDepth$. Near the dam, distributions reach values of over $30 \ \unit{m}$ for some extreme values, which agrees with observations \citep{Robertson2019}. The development of $\maxFlowDepth$ presents local increases and decreases instead of a monotonically decrease, probably associated with topographic-related aspects, e.g. accumulation zones, lateral spreading and changes in the valley geometry. Uncertainty is heavily non-uniform throughout the tailings main flow path. At points near the dam (P1 - P4), significant variability is explained by hydrograph-related parameters, where $\totalTimeHyd$ controls magnitudes of flow rate and the hydrograph's duration, and $\relativeTimePeak$ is associated to the timing of peak flow rate. Then, at points P5 - P11, variability tends to be lower and completely related to $\yieldStress$, according to the Sobol' indices. Thus, a transition is observed from P4 to P5, where hydrograph-related parameters stop being dominant and yield stress takes over as the main influence to maximum flow depth variability. Two factors explain this transition. Firstly, due to the relatively short characteristic time of hydrograph, its effect is limited to the region near the dam (P1 to P4). Secondly, after P5, the terrain is shaped as a narrow valley, favouring the effects of accumulation brought by $\yieldStress$. Finally, after point P11, significant variability is observed because that is a region where the tailings flow is susceptible to arresting until stoppage. Lower whiskers and outliers of distributions approach zero or lower values of maximum flow depth. At the same time, many scenarios have sufficient momentum to maintain the flow with significant depth, which explains the remaining values of the distribution. Physically, $\yieldStress$ controls the flow runout: a scenario with lower yield stress presents a higher runout value; conversely, higher yield stress results in lower runout. This behaviour is confirmed by the Sobol' indices, with clear dominance of yield stress.  

\begin{figure}[H]
    \centering
    \includegraphics[width=0.9\linewidth]{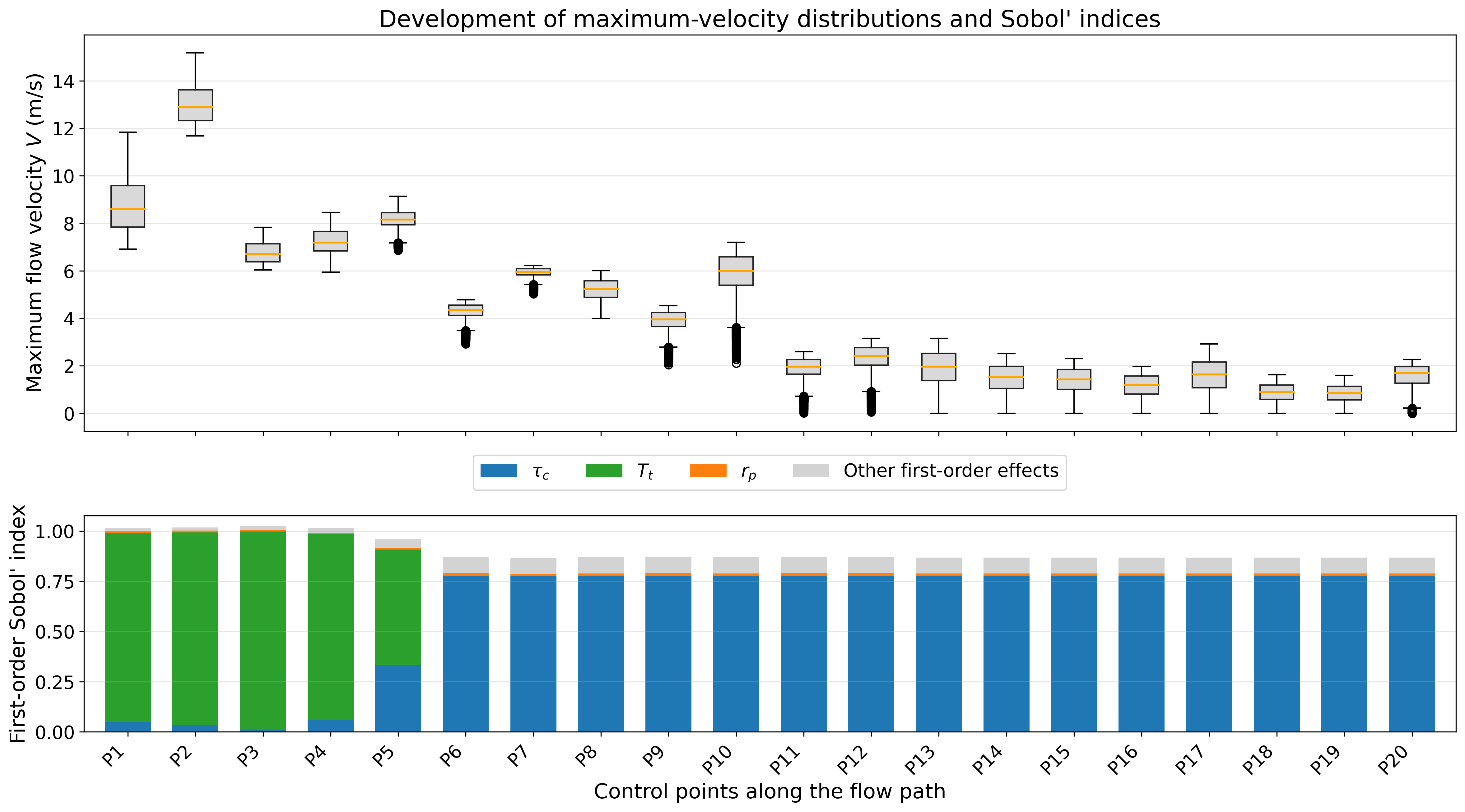}
    \caption{Spatial development of $\maxVel$ over 20 points along the main flow path. Distributions of data and correspondent Sobol' indices are shown in the top and bottom panels, respectively. $10000$ input scenarios were used to generate the output distributions. PF samples of size $10000$ were used for Sobol' indices estimation.}
    \label{fig:spatial_development_V}
\end{figure}

Figure \ref{fig:spatial_development_V} shows similar patterns and observations for $\maxVel$ in comparison to the spatial development of $\maxFlowDepth$. Higher maximum velocities (from $6$ to $14 \ \unit{m s^{-1}}$) indicate a rapid flow near the dam. On the other hand, distant points tend to show lower maximum velocity (order of $2 \ \unit{m s^{-1}}$) since the terrain does not present significant slopes and, therefore, friction decelerates the flow. Uncertainty is also non-uniform. Near the dam, maximum velocity is sensitive to $\totalTimeHyd$, which controls the magnitude of flow rate; variable $\relativeTimePeak$ is non-influential since the timing is irrelevant for the maximum flow velocity. Around point P5 to P6, variable $\yieldStress$ starts to be dominant, as observed in Fig. \ref{fig:spatial_development_H}. In this case, other effects, such as interactions and volumetric concentration, are more significant in the analysis of maximum velocity for distant points.

\begin{figure}[H]
    \centering
    \includegraphics[width=0.9\linewidth]{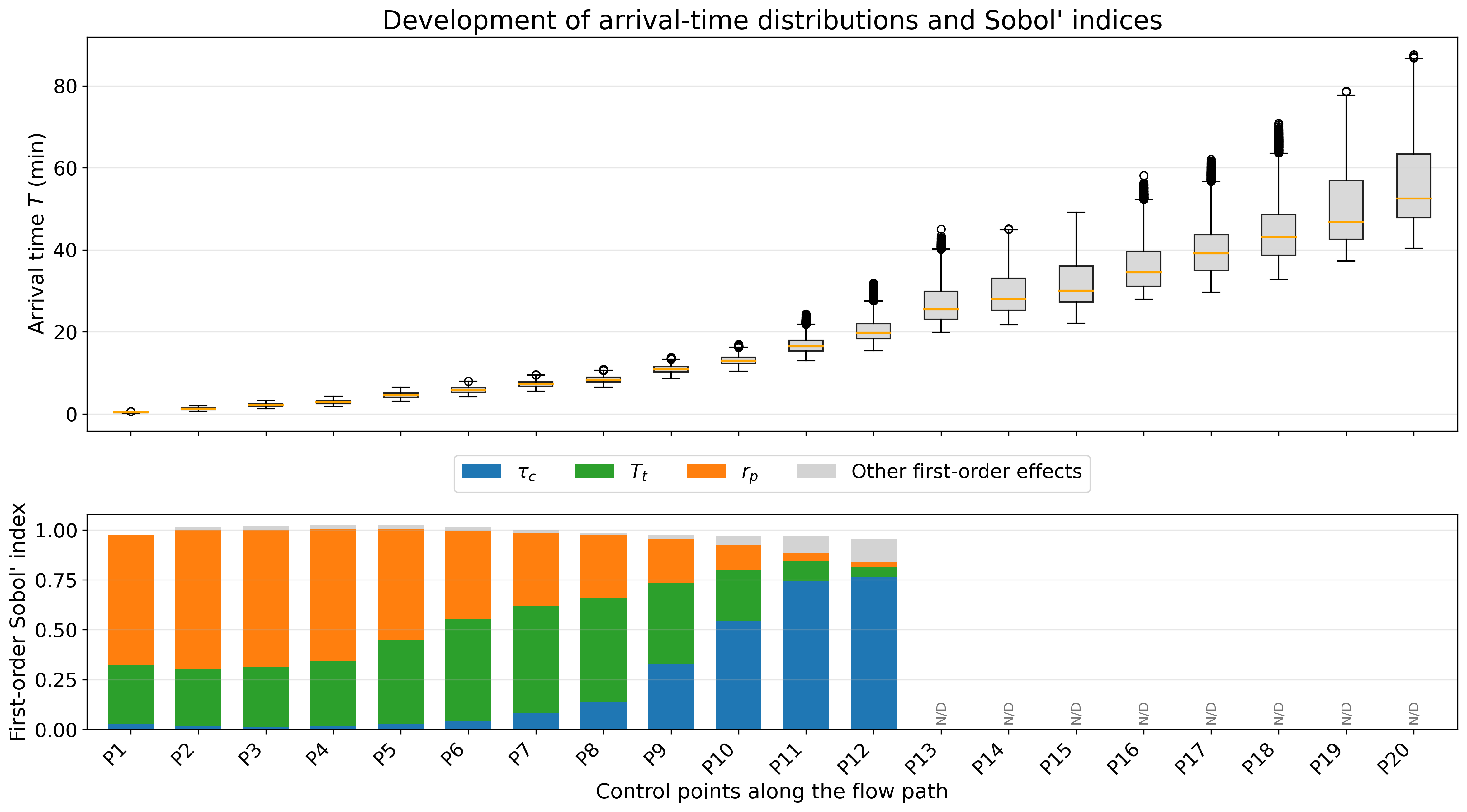}
    \caption{Spatial development of $\arrivalTime$ over 20 points along the main flow path. Distributions of data and correspondent Sobol' indices are shown in the top and bottom panels, respectively. $5000$ input scenarios were used to generate the output distributions. PF samples of size $5000$ were used for Sobol' indices estimation.}
    \label{fig:spatial_development_T}
\end{figure}

Figure \ref{fig:spatial_development_T} presents the spatial development of arrival times. Contrary to the spatial development of previous outputs, this case shows an expected pattern that increases monotonically, for both the associated uncertainty and median values. Points near the dam are reached almost immediately after the failure. Therein, the variability is governed primarily by the $\relativeTimePeak$, followed by $\totalTimeHyd$. Differently from the previous analyses of $\maxFlowDepth$ and $\maxVel$, the influence of hydrograph-related parameters persists until around P11, where $\yieldStress$ starts to dominate. Physically, $\yieldStress$ controls how fast the tailings flow propagate, affecting significantly the arrival times and, therefore, their variability.

\begin{figure}[H]
    \centering
    \includegraphics[width=0.9\linewidth]{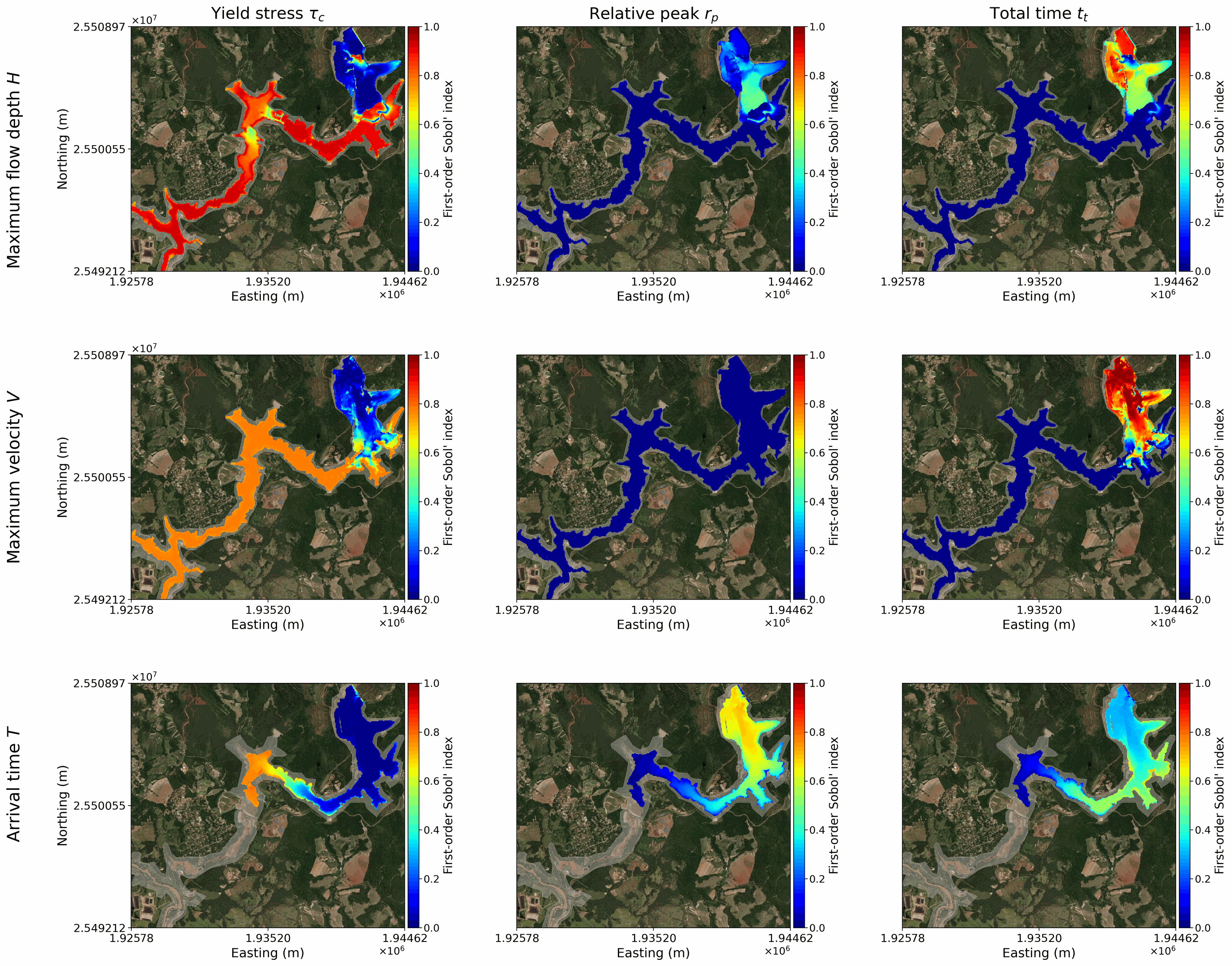}
    \caption{Panel containing first-order Sobol' maps, where each row and column represent a hazard-related output and an input variable, respectively. Hot and cold colors represent high and low influence, respectively.}
    \label{fig:spatial_sobol}
\end{figure}

In fact, the first-order Sobol' indices can be extended to the entire domain, as shown by Fig. \ref{fig:spatial_sobol}. It confirms the non-uniformity of Sobol' indices shown by Figs. \ref{fig:spatial_development_H}, \ref{fig:spatial_development_V} and \ref{fig:spatial_development_T}, indicating clearly the areas controlled by hydrograph-related parameters and by $\yieldStress$. The sensitivity maps also confirms the similarity of behaviour of Sobol' indices between maximum flow depth and maximum velocity, while detailing the extended influence of $\relativeTimePeak$ and $\totalTimeHyd$ over the arrival times variability. Physically, the sensitivity maps confirm the importance of hydrograph-parameters, i.e. breach mechanism, over the regions near the failure, whereas rheological properties, specially $\yieldStress$, control the distant areas of the tailings flow. In short, both sources of uncertainty are important in different areas of the tailings flow. 

Very low influence of the remaining input variables must also be addressed. In the case of released volume, it is usually a very important variable to the general hydrodynamics of the flow, controlling the hydrograph propagated to the domain; in terms of sensitivity analysis, this was shown conclusively by \cite{Ghahramani2024}. However, in this work, the statistical support is narrow due to the accurate estimations made by literature (see Tab. \ref{tab:PDFs}), resulting in low uncertainty of input scenarios and, therefore, low uncertainty propagated to the outputs. Concerning volumetric concentration, its role is to control the density of the bulk fluid, since rheological parameterization is independent of $\volumetricConc$ in this work. It means that $\volumetricConc$ is limited to scale momentum balance and resistance terms of SWE system, weakening its overall influence on the flow. Regarding plastic viscosity, \cite{Ghahramani2024} and \cite{Melo2022} have also found that plastic viscosity influence is negligible. The role of plastic viscosity in SWE context is to compose the basal shear stress, which is essentially the same as yield stress and friction due to land roughness. Usually, plastic viscosity presents nominal values one order of magnitude smaller in comparison to yield stress for tailings rheology and, in addition, it is scaled with shear rate. Therefore, plastic viscosity contribution tend to be low in comparison with the other components of basal shear stress.

\subsection{Analyses and comparisons with literature at control points}
\label{subsec:control_points}

\begin{figure}[H]
    \centering
    \includegraphics[width=0.8\linewidth]{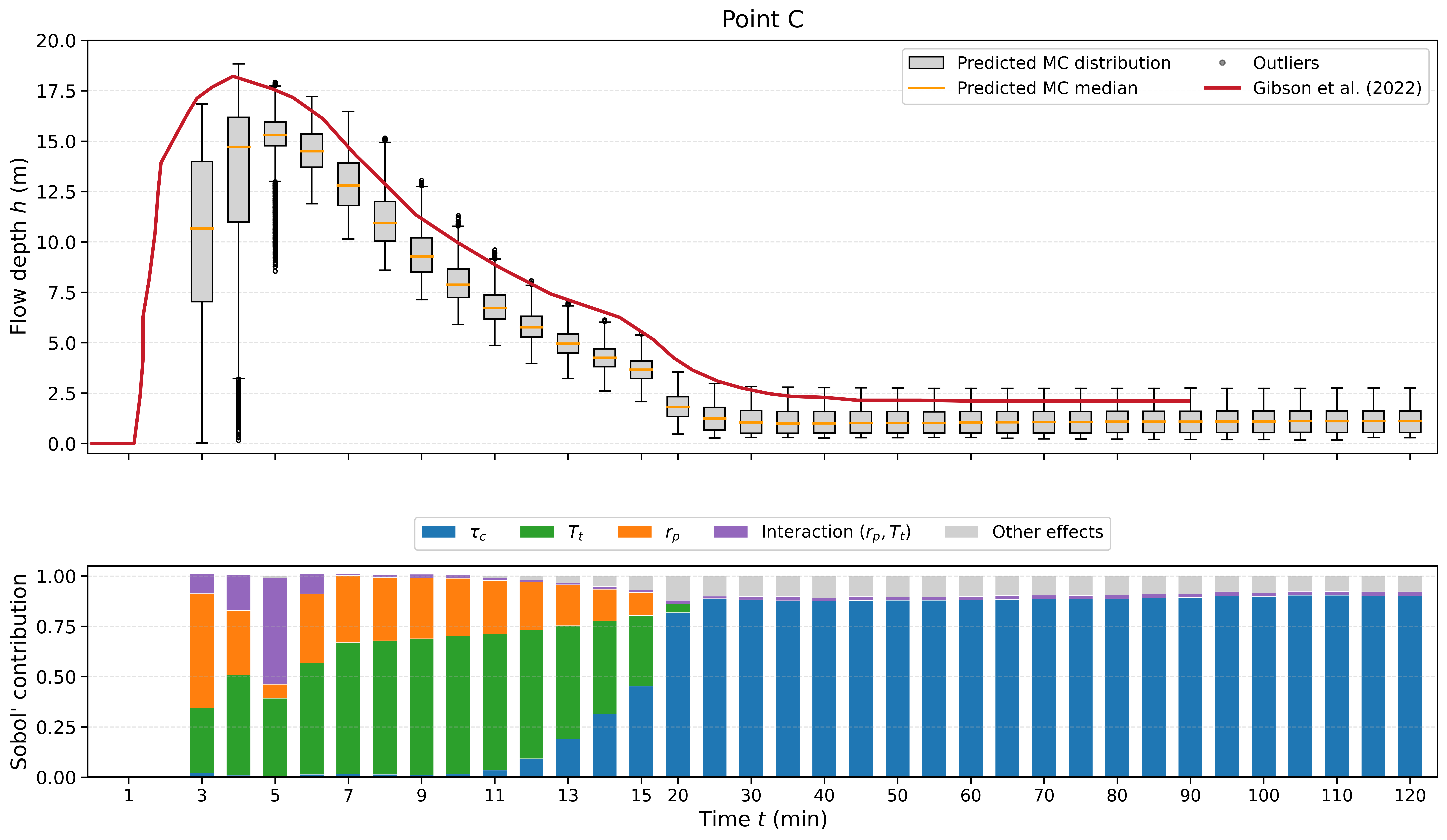}
    \caption{Temporal development of flow-depth distributions at control point C (top panel). Boxplots are used to represent the distribution, where the median is represented by orange segment. Data from literature is represented by a red continuous line. Sensitivity indices for each time instant are indicated in the bottom panel through stacked bars.}
    \label{fig:flow_depth_over_time_POINT_C}
\end{figure}

Figure \ref{fig:flow_depth_over_time_POINT_C} shows the temporal development of flow-depth distributions and the corresponding Sobol' indices at control point C. The point is affected very early, approximately between $3$ and $4 \ \unit{min}$ after failure; flow depth increases sharply until reaching a median maximum of about $15 \ \unit{m}$ around $5 \ \unit{min}$. This rapid increase is associated with the arrival of the peak flow rate. After the peak, flow depth decreases more gradually and reaches a quasi-stabilized final condition after approximately $25 \ \unit{min}$, with depths mostly between $0.5$ and $3.5 \ \unit{m}$. The deterministic result reported by \cite{Gibson2022} presents a similar temporal behavior, although it is generally located in the upper part of the simulated distributions, which may be related to differences in Manning roughness assumptions and hydrograph shape.

The Sobol' indices indicate that the uncertainty at point C is initially governed by hydrograph-related parameters, especially $\relativeTimePeak$, $\totalTimeHyd$, and their interaction. During the rising phase, $\relativeTimePeak$ controls the timing of the peak release, while $\totalTimeHyd$ controls the overall release duration and the magnitude of the imposed flow rate. The interaction between both variables becomes especially relevant near the maximum flow depth, indicating that the effect of an early or late peak depends on the total duration of the release. After the peak, the influence of $\relativeTimePeak$ progressively vanishes, while $\yieldStress$ becomes increasingly dominant during the recession and final-deposition phases. This transition indicates that, once the hydrograph effects have passed, the remaining uncertainty in flow depth is mainly controlled by the viscoplastic resistance of the material.

\begin{figure}[H]
    \centering
    \includegraphics[width=0.8\linewidth]{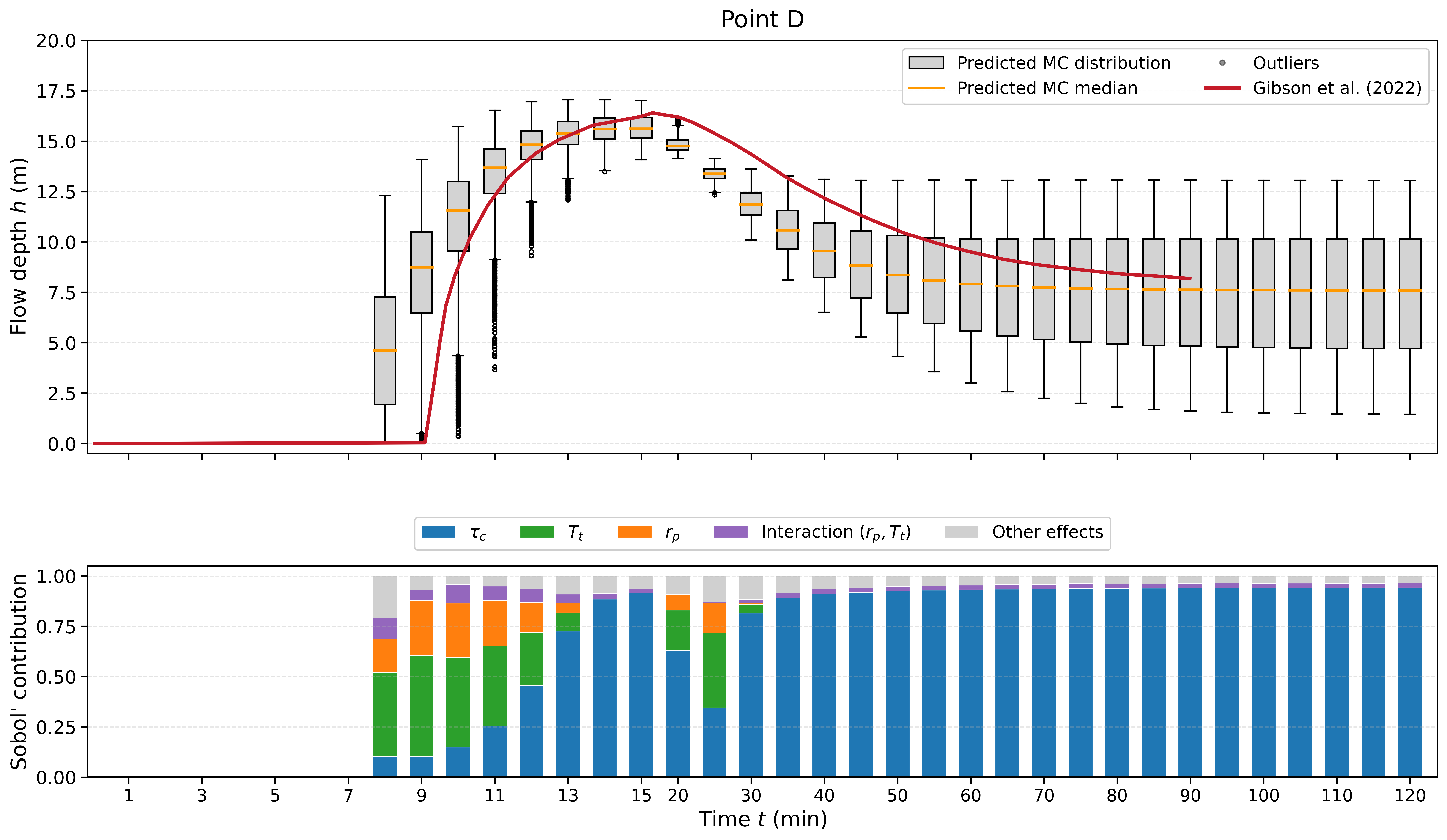}
    \caption{Temporal development of flow-depth distributions at control point D (top panel). Boxplots are used to represent the distribution, where the median is represented by orange segment. Data from literature is represented by a red continuous line. Sensitivity indices for each time instant are indicated in the bottom panel through stacked bars.}
    \label{fig:flow_depth_over_time_POINT_D}
\end{figure}

Figure \ref{fig:flow_depth_over_time_POINT_D} presents the same analysis for point D. The flow reaches this location later, around $8$ to $9 \ \unit{min}$, and develops toward a maximum median depth of approximately $15 \ \unit{m}$ between $12$ and $15 \ \unit{min}$. Compared with point C, the rise is less abrupt and the recession phase is considerably longer, with median values stabilizing only after approximately $60 \ \unit{min}$. Final depths are also much larger than at point C, which is consistent with the local terrain configuration: point D is located in a narrower valley that favors channelization and accumulation, whereas point C is more exposed to lateral spreading. The deterministic solution from \cite{Gibson2022} follows a similar trend, with a slight delay in arrival and comparable depths around the peak.

At point D, uncertainty is again strongly time-dependent. The rising phase is influenced by hydrograph-related variables, but $\yieldStress$ becomes important earlier than at point C and dominates after the peak. A brief reappearance of hydrograph-related effects during the recession may be associated with the remaining upstream volume and the local valley configuration. In the quasi-final and final-deposition phases, the flow-depth distributions remain wide, with values ranging approximately from $1.5$ to $13.5 \ \unit{m}$. This uncertainty is almost entirely explained by $\yieldStress$, indicating that the final accumulation at point D is mainly controlled by the resistance of the tailings flow and its ability to continue flowing downstream.

\begin{figure}[H]
    \centering
    \includegraphics[width=0.8\linewidth]{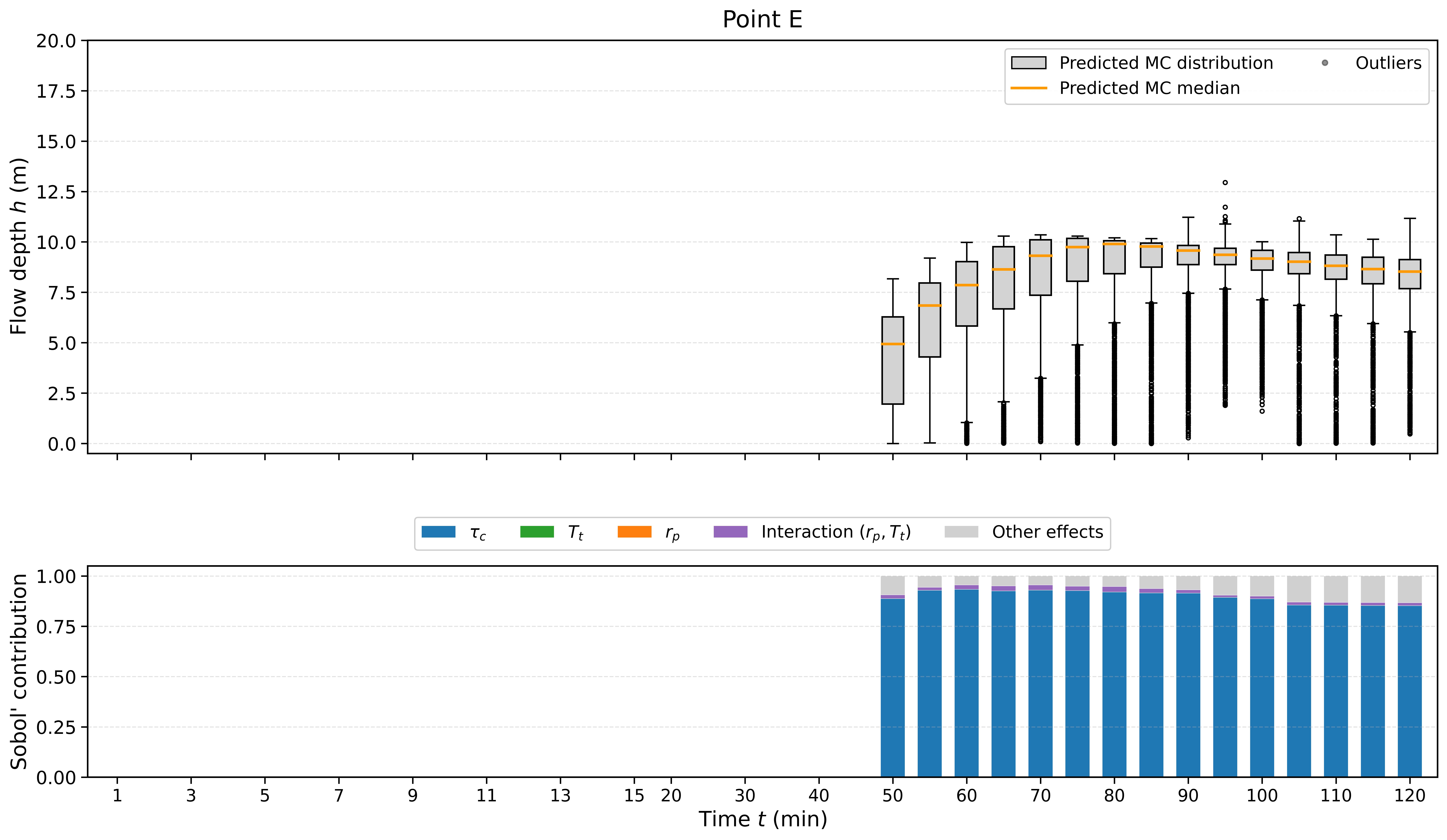}
    \caption{Temporal development of flow-depth distributions at control point E (top panel). Boxplots are used to represent the distribution, where the median is represented by orange segment. Sensitivity indices for each time instant are indicated in the bottom panel through stacked bars.}
    \label{fig:flow_depth_over_time_POINT_E}
\end{figure}

Finally, Fig. \ref{fig:flow_depth_over_time_POINT_E} shows the temporal response at point E. Because this point is located much farther downstream, around $10 \ \unit{km}$ from Dam B1, the flow arrives much later, between approximately $50$ and $115 \ \unit{min}$. The increase in flow depth is slower than at points C and D, requiring about $30 \ \unit{min}$ to reach maximum median depths close to $10 \ \unit{m}$. No literature data was found for the temporal development of flow depth for this point. The distributions are also markedly asymmetric, with a compressed upper half and a wider lower tail. This suggests that scenarios with sufficient mobility, i.e. low yield stress, produce depths constrained by the local terrain, whereas scenarios closer to stoppage generate shallower flows and form the lower part of the distribution. The Sobol' indices at point E show that hydrograph-related parameters have little influence on flow-depth uncertainty. By the time the flow reaches this distant location, the initial hydrograph effect has been strongly attenuated by frictional resistance and spreading. Consequently, uncertainty is controlled almost exclusively by $\yieldStress$.

\begin{figure}[H]
    \centering
    \includegraphics[width=0.95\linewidth]{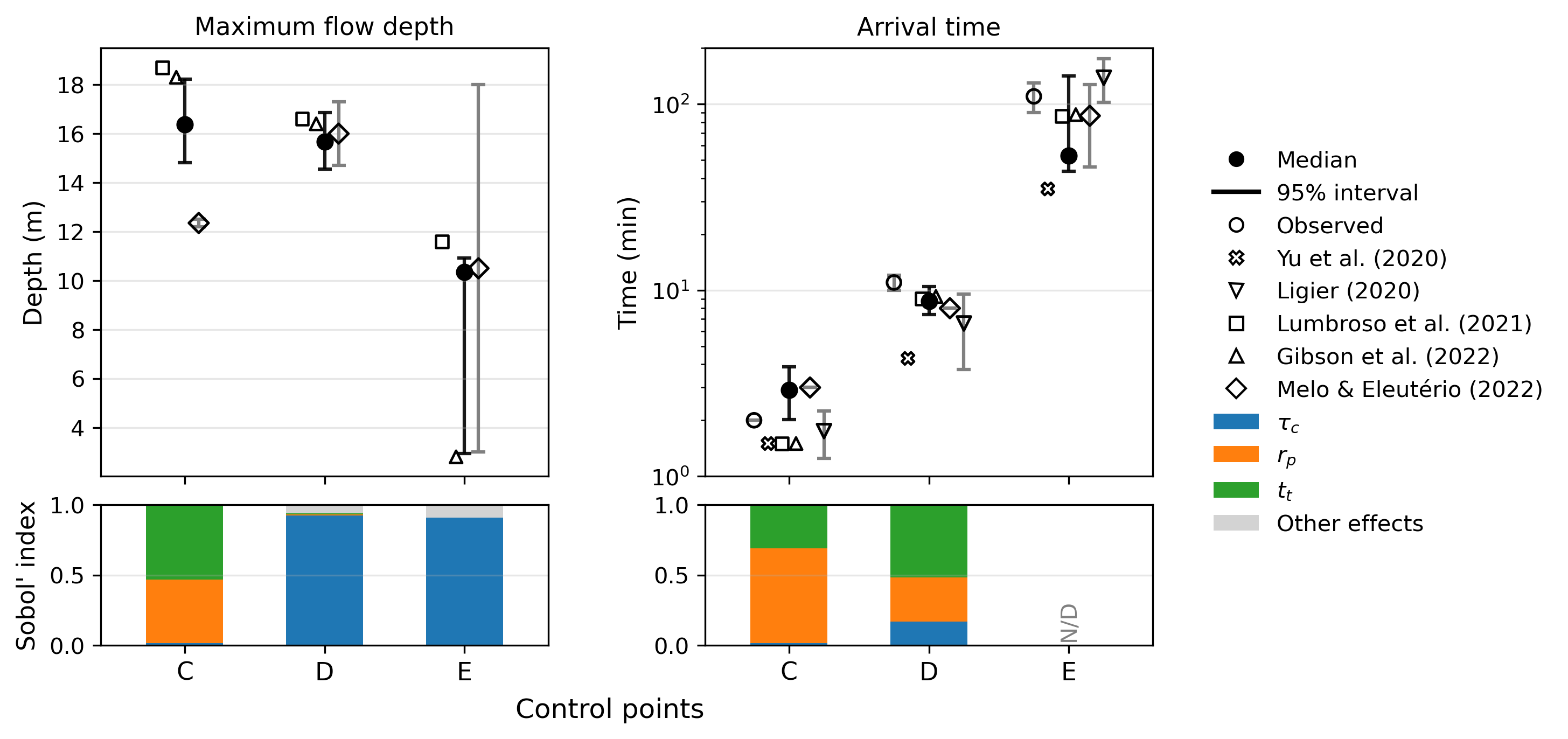}
    \caption{Comparison of hazard-related properties data with literature, at control points C, D and E. Top panel presents maximum flow depth (left) and arrival times (right). First-order Sobol' indices of the respective points and properties are defined in the bottom panel.}
    \label{fig:pointwise_uncertainty_sobol}
\end{figure}

As mentioned in Tab. \ref{tab:literature_results}, literature focuses the analyses at three control points. They are referred as points C, D and E, corresponding to points P4, P8 and P20 defined for this work. Figure \ref{fig:pointwise_uncertainty_sobol} compares the probabilistic results obtained at control points C, D and E with values reported in previous Brumadinho dam-breach studies. In general, the literature estimates are located within or close to the $95\%$ intervals obtained in the present work, indicating broad consistency between the probabilistic simulations and previous deterministic or observational estimates. 


\section{Conclusions}  

This work applied a probabilistic approach to study the tailings dam-breach flow caused by the rupture of Dam B1 in Brumadinho on 2019. The study used the deterministic solver HEC-RAS v6.6 to compute hazard-related hydrodynamical properties at control points, e.g. flow depth, and at the entire domain, e.g. maximum flow depth, maximum velocity, arrival times and flooding probability. Contrary to a purely deterministic approach, which relies on few representative scenarios, the proposed probabilistic approach allowed to forward-propagate relevant input uncertainties associated to physical properties of tailings, hydrograph modelling and released volume. As result, output properties were described as distributions, reflecting the variability caused by input uncertainty and providing a more detailed description of the flow behaviour under effects of uncertainties. Furthermore, local and spatially-distributed sensitivity indices were used to explain which input variables were most influential to the variability of outputs. From a methodological perspective, both the isolated usage of Gaussian process regression (GPR) and its combination with principal component analysis (PCA) for dimensionality-reduction were successful. For properties affected by temporal discontinuities of flooded/non-flooded states, e.g. flow depth and arrival times, random forest classifiers (RFCs) were able to classify the flooded/non-flooded states efficiently, filtering data for a more accurate metamodeling representation by GPR while preserving physical consistency.

In general, output uncertainty is non-uniform and is influenced by different input variables in both space and time, quantified by the Sobol' indices. Two main regions with different regimes were identified. Near the dam, hydrograph-related parameters govern output variability, whereas resistance effects are unable to counterbalance effectively the flow inertia. Farther downstream, yield stress is dominant. Its effect attenuate the momentum of the flow, causes accumulation of volume in certain regions and eventually arrests the flow to stoppage. The accumulation, and therefore the influence of yield stress, is particularly notable in narrow valleys. At late times and distant locations, the influence of hydrograph is completely dissipated. Data distributions obtained by the probabilistic approach are consistent with previous deterministic back-analyses of the Brumadinho dam-breach flow. Some deviations are present due to different modelling assumptions, e.g. distribution of Manning coefficients, hydrograph shape and deterministic methods to solve the governing equations.

Some limitations and future improvements should be acknowledged. The triangular hydrograph represents a simplified representation of a complex mobilization phenomenon, adopted to describe the general dynamics of the process with few interpretable parameters. Other hydrograph models could be used, specially for a more detailed evaluation of the initial dynamics of the flow in regions near the dam. The deterministic solver also presents limitations and was chosen parsimoniously to provide large-scale properties. Shallow-water approximation and one-phase fluid approach cannot represent all processes involved, e.g. material heterogeneity, entrainment, deposition and three-dimensional effects. Some models, specially those with a more geotechnical approach, can be useful to provide more detail information, although a balance of computational cost and accuracy should be found to apply a probabilistic approach. Finally, although the metamodeling strategy enabled prediction of maps and distributions, its accuracy depends on the training design, the deterministic simulations and the treatment of flooded/non-flooded states.

Overall, this study demonstrates that the proposed probabilistic approach coupled with GSA can substantially enrich tailings dam-breach assessments, specially for an engineering point-of-view. It provides information that is not accessible from a small number of deterministic simulations, e.g. flooding probability, uncertainty bands and sensitivity indices. In addition to identify where resources should be allocated to reduce predictive uncertainty through sensitivity indices, the results can directly support risk-informed tailings dam-breach studies under incomplete knowledge.

\appendix

\section{Prediction of hazard-related output maps}
\label{app:algorithm}

Algorithm~\ref{algo:prediction_intro} summarizes the PCA--GPR regression workflow used to predict the maximum flow depth, maximum flow velocity and mean-imputed arrival-time maps. The procedure comprises an offline phase, in which the PCA representation and the GPR metamodels are constructed from the available HEC-RAS simulations, and an online phase, in which these metamodels are used to predict a new high-dimensional map at low computational cost compared with an additional hydraulic simulation.

\begin{algorithm}[H]
\caption{Offline construction and online prediction of a high-dimensional map using PCA and GPR.}
\label{algo:prediction_intro}
\begin{algorithmic}[1]

\STATE \textbf{Offline input:} DoE $\DoE$; training data $\dataMatrixPCA$; number of retained basis components $\nbasis$.

\STATE \textbf{Offline phase:}

\STATE Compute the dimension-wise mean $\dimensionMean$ with Eq. \ref{Eq:dimension_wise_mean} and center the training data, obtaining $\centeredData$.

\STATE Calculate the empirical covariance matrix $\covarianceMatrixPCA$ with Eq. \ref{Eq:covariance_matrix_PCA}.

\STATE Solve the eigenvalue problem
$\covarianceMatrixPCA \basis_{\indBasis}
= \lambda_{\indBasis}\basis_{\indBasis}$.

\STATE Sort the eigenvalues
$\lambda_1 \geq \dots \geq \lambda_{\nOutputDimensions}$
and the associated PCA basis components.

\STATE Retain the first $\nbasis$ PCA basis components and define the truncated PCA basis $\basisMatrixPCA$.

\STATE Project the centered data onto the PCA basis:
$\coefficientsMatrixPCA
= \centeredData \basisMatrixPCA^{\top}$.

\FOR{$\indBasis = 1$ \TO $\nbasis$}
    \STATE Define
    $\coefficientsMatrixPCA_{.,\indBasis}
    =
    \left(
    \coef_{\indBasis}
    \left(
    \inputPoint^{(\indPoints)}
    \right)
    \right)_{\indPoints=1,\dots,\nDoE}$.
    
    \STATE Train the $\indBasis$-th GPR metamodel
    $\GaussianProcess_{n}^{(\indBasis)}$
    using $\DoE$ and
    $\coefficientsMatrixPCA_{.,\indBasis}$.
\ENDFOR

\STATE \textbf{Online input:} New input scenario $\inputPoint^{*}$.

\STATE \textbf{Online phase:}

\FOR{$\indBasis = 1$ \TO $\nbasis$}
    \STATE Predict
    $\widehat{\coef}_{\indBasis}(\inputPoint^{*})$
    using
    $\GaussianProcess_{n}^{(\indBasis)}$.
\ENDFOR

\STATE Define the predicted basis-coefficient vector:
\[
\widehat{\coef}(\inputPoint^{*})
=
\left(
\widehat{\coef}_{1}(\inputPoint^{*}),
\dots,
\widehat{\coef}_{\nbasis}(\inputPoint^{*})
\right)^{\top}.
\]

\STATE Reconstruct the predicted high-dimensional map:
\[
\widehat{\outputPoint}(\inputPoint^{*})
=
\dimensionMean
+
\basisMatrixPCA^{\top}
\widehat{\coef}(\inputPoint^{*}).
\]

\STATE \textbf{return}
$\widehat{\outputPoint}(\inputPoint^{*})$.

\end{algorithmic}
\end{algorithm}

For arrival time, Algorithm~\ref{algo:prediction_intro} produces the continuous mean-imputed map predicted by the PCA--GPR regression component. The RFC-based flooded/non-flooded mask is subsequently applied to this prediction to recover the physically consistent arrival-time map, as described in Subsection~\ref{subsec:treatment_outputs}.

\section{Validation of predictions}
\label{app:validation}

Some representative validations of the PCA-GPR methodology are shown here; please check the Supplementary Material for all PCA-GPR methodology validation results and for the RFC validation. Here, $\nDoE_{val}$ is the number of validation scenarios, $y_{\indOutputs}^{(i)}$ and $\hat{y}_{\indOutputs}^{(i)}$ are the reference and PCA--GPR-predicted values, respectively, at output dimension $\indOutputs$ for validation scenario $i$, and $\bar{y}_{\indOutputs}$ is the corresponding mean of the reference validation values. The main metrics to validate the prediction PCA-GPR methodology are the Nash-Sutcliffe efficiency ($Q^2$ metric): 

\begin{equation}
\label{eq:Q2}
    Q^2_{\indOutputs} = 1 - \frac{\sum_{i=1}^{\nDoE_{val}}\left(y_\indOutputs^{(i)} - \hat{y}_{\indOutputs}^{(i)}\right)^2}{\sum_{i=1}^{\nDoE_{val}}\left(y_\indOutputs^{(i)} - \bar{y}_{\indOutputs}\right)^2}
\end{equation}

and the root mean squared error ($\mathrm{RMSE}$):

\begin{equation}
\label{eq:RMSE}
    \mathrm{RMSE}_{\indOutputs} = \sqrt{\frac{1}{\nDoE_{val}} \sum^{\nDoE_{val}}_{i=1} \left( y_{\indOutputs}^{(i)} - \hat{y}_{\indOutputs}^{(i)} \right)^2}
\end{equation}

where $\nDoE_{val}$ is the size of the validation set, $y_\indOutputs^{(i)}$ is the reference value for validation scenario $i$, and $\hat{y}_\indOutputs^{(i)}$ is its PCA-GPR prediction. A value of $Q^2$ close to one indicates that the prediction reproduces most of the validation variance, whereas $\mathrm{RMSE}_{\indOutputs}$ quantifies the prediction error in the physical unit of the output. The $Q^2$ metric is left undefined at output dimensions with zero validation variance.

\begin{figure}[H]
\centering
\includegraphics[width=0.9\linewidth]{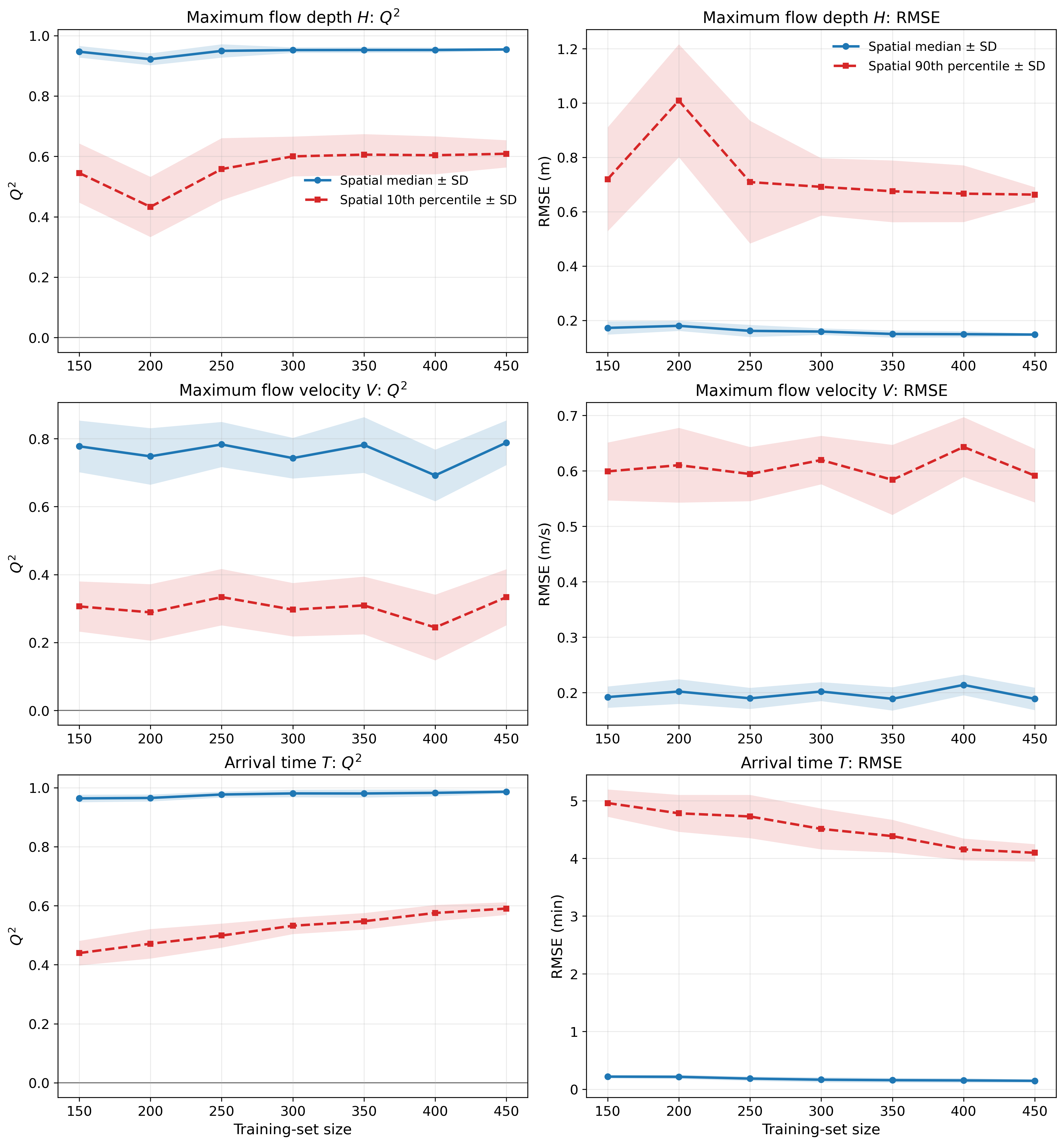}
\caption{Repeated learning curves for the PCA--GPR predictions. The spatial median and adverse percentile are reported for $Q^2$ and $\mathrm{RMSE}$ as functions of the training-set size. The central curves represent the median across ten random training/validation partitions, and the shaded regions represent one standard deviation across these repetitions. The validation-set size was fixed at $100$ scenarios. Rows correspond to maximum flow depth (top), maximum flow velocity (middle), and arrival time (bottom); columns correspond to $Q^2$ (left) and $\mathrm{RMSE}$ (right).}
\label{fig:learning_curve_training_size}
\end{figure}

The influence of the size of the training set was assessed through a repeated random holdout procedure, generating learning curves. Training sets containing between $150$ and $450$ scenarios were considered, while the size of the validation set was fixed at $100$ scenarios. For every repetition, $Q^2$ and $\mathrm{RMSE}$ were calculated at each eligible output dimension. The resulting spatial fields were summarized using their median and an adverse percentile: the 10th percentile for $Q^2$ and the 90th percentile for $\mathrm{RMSE}$. The spatial median represents the prediction performance over a typical mesh cell, whereas the adverse percentiles characterize the less accurately predicted portion of the domain without being controlled by isolated extreme values. For example, the 10th percentile of $Q^2$ is exceeded by approximately 90\% of the evaluated cells, while the 90th percentile of $\mathrm{RMSE}$ is exceeded by approximately 10\%. The central curves in Fig.~\ref{fig:learning_curve_training_size} represent the median of these spatial summaries across the ten repetitions, and the shaded regions represent their standard deviation. A common eligible spatial support was used across the training-set sizes for each output to ensure comparability.

Overall, the improvements between $400$ and $450$ scenarios are small for the median metrics and remain modest for the adverse percentiles. The use of $450$ training scenarios therefore provides the best performance among the evaluated configurations while preserving an independent validation set of $100$ scenarios. The curves indicate stabilization for maximum flow depth and no systematic sample-size dependence for maximum flow velocity, although the adverse-percentile arrival-time metrics suggest that some localized predictions could still benefit from additional training data.

\begin{figure}[H]
    \centering

    \includegraphics[width=0.95\linewidth]{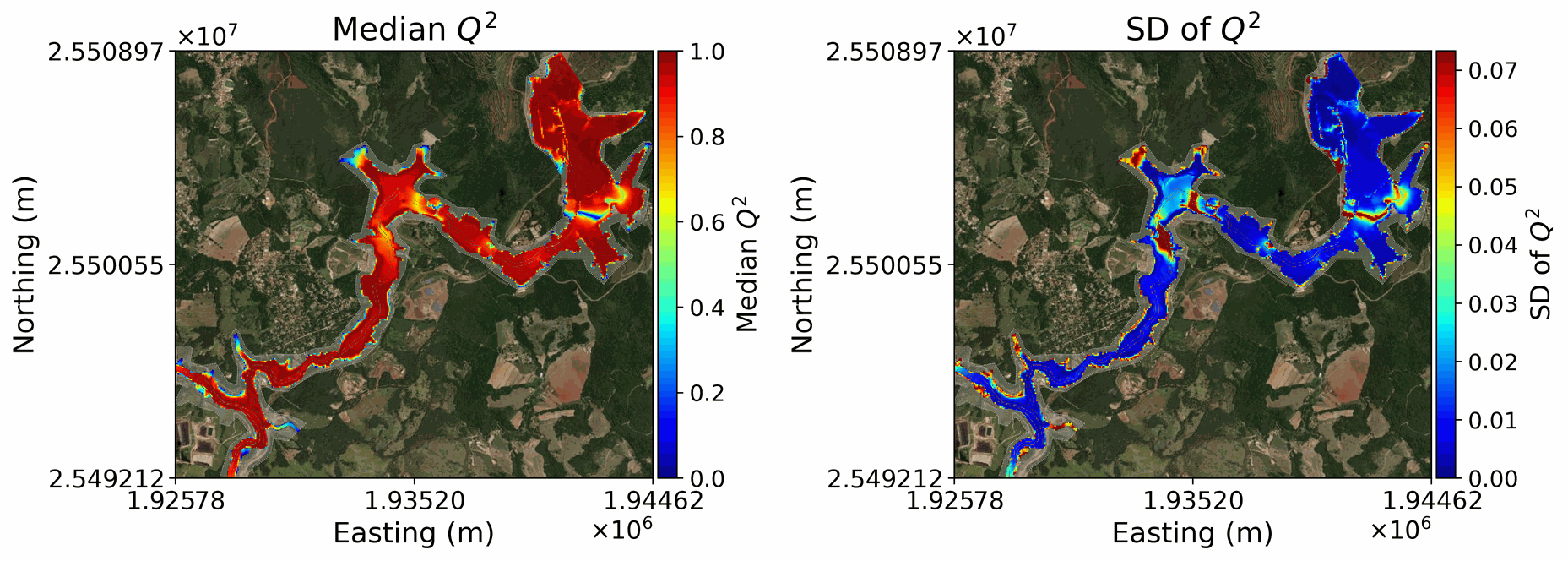}

    \vspace{0.5cm}

    \includegraphics[width=0.95\linewidth]{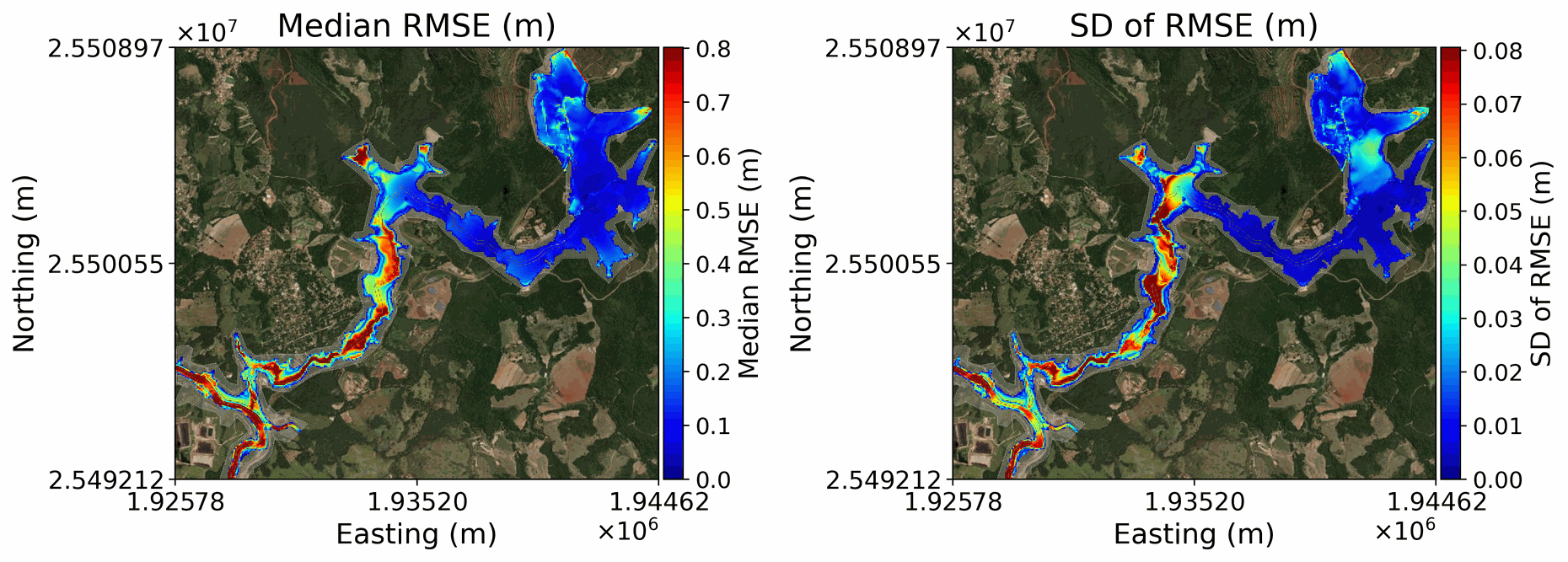}

    \caption{Spatial validation of the maximum-flow-depth PCA--GPR predictions obtained from ten repeated random holdout partitions with $450$ training and $100$ validation scenarios. Upper panels: median (left) and standard deviation (right) of $Q^2$. Lower panels: median (left) and standard deviation (right) of $\mathrm{RMSE}$.}
    \label{fig:q2_rmse_maxdepth}
\end{figure}

A repeated random holdout validation with $450$ training and $100$ validation scenarios was additionally used to evaluate the spatial distribution and split-to-split stability of the prediction metrics. The procedure was repeated ten times with independently randomized partitions. For each mesh cell, the median metric represents its typical validation performance, while the standard deviation measures the sensitivity of that performance to the selection of training and validation scenarios. Figure \ref{fig:q2_rmse_maxdepth} shows the spatially distributed median and standard deviation of $Q^2$ and $\mathrm{RMSE}$ for the maximum flow depth, obtained from the repeated random holdout validation procedure, as described in item \ref{subsubsec:GPR}.

The median $Q^2$ is close to one over most of the modeled inundated domain, showing that the PCA--GPR predictions reproduce a large fraction of the local flow-depth variance. Lower values are concentrated in limited regions, particularly along narrow flow corridors, confluences, and parts of the inundation boundary. The standard deviation of $Q^2$ is generally low, thus the spatial distribution of $Q^2$ is largely stable with respect to the random selection of training and validation scenarios. The median $\mathrm{RMSE}$ is below approximately $0.2$~m over much of the domain, while localized values between approximately $0.5$ and $0.8$~m occur mainly along narrow flow paths and regions of greater flow-depth variability.

\printcredits

\section*{Declaration of competing interests}
The contact author has declared that none of the authors has any competing interests.

\section*{Data availability}
Data will be made available on request.

\section*{Acknowledgements}
The authors gratefully acknowledge Olivier Roustant (Institut National des Sciences Appliquées de Toulouse – INSA Toulouse, Department of Mathematics and Applications) for his teaching, guidance and insightful discussions on the probabilistic aspects of this work. This research was supported by the Fundação de Amparo à Pesquisa do Estado de São Paulo (FAPESP) through Ph.D. scholarship grants 2022/05184-1 and 2023/08472-0. This study was also partly financed by the Coordenação de Aperfeiçoamento de Pessoal de Nível Superior – Brasil (CAPES) – Finance Code 001. The Article Processing Charge (APC) for the publication of this research was funded by the Coordenação de Aperfeiçoamento de Pessoal de Nível Superior – Brasil (CAPES) (ROR identifier: 00x0ma614).

\bibliographystyle{cas-model2-names}

\bibliography{cas-refs}



\end{document}